\documentclass[aps,groupedaddress,superscriptaddress,showpacs,prb,floatfix]{revtex4}
\usepackage{graphicx,rotating,subfigure,amsmath,amsfonts,amssymb,delarray}

\usepackage{color}

\usepackage{graphicx}
\usepackage{dcolumn}
\usepackage{bm}
\usepackage{setspace}

\newcommand{\be}{\begin{eqnarray}}
\newcommand{\ee}{\end{eqnarray}}
\def\refeq#1{(\ref{#1})}
\def\d{\mbox d}

\def\nn{\nonumber}
\def\i{\int_{-\infty}^{\infty}}

\def\Or{\mathcal O}
\def\g{\gamma}

\def\ve{\varepsilon}

\def\l{\left}
\def\r{\right}
\def\te{\mbox{e}}
\def\rmi{{\rm i}}

\begin{document}
\bibliographystyle{apsrev}
\title{Lattice {\em vs.} continuum theory of the periodic Heisenberg chain}
\author{Michael Bortz}
\email{bortz@physik.uni-kl.de}
\affiliation{Technische Universit\"at Kaiserslautern, Fachbereich Physik, Erwin-Schr\"odinger-Str., D-67663 Kaiserslautern, Germany}
\author{Michael Karbach}
\affiliation{Bergische Universit\"at Wuppertal, Fachbereich Mathematik und Naturwissenschaften, Physik, D-42097 Wuppertal, Germany}
\author{Imke Schneider}
\affiliation{Technische Universit\"at Kaiserslautern, Fachbereich Physik, Erwin-Schr\"odinger-Str., D-67663 Kaiserslautern, Germany}
\author{Sebastian Eggert}
\affiliation{Technische Universit\"at Kaiserslautern, Fachbereich Physik, Erwin-Schr\"odinger-Str., D-67663 Kaiserslautern, Germany}
\date{\today}
\begin{abstract}
We consider the detailed structure of low energy excitations in the periodic spin-1/2 $XXZ$ Heisenberg chain. By performing a perturbative calculation of the non-linear corrections to the Gaussian model, we determine the exact coefficients of asymptotic expansions in inverse powers of the system length $N$ for a large number of low-lying excited energy levels. This allows us to calculate eigenenergies of the lattice model
up to order $\mathcal O (N^{-4})$, without having to solve the Bethe Ansatz equations. 
At the same time, it is possible to express the exact eigenstates of the lattice model in terms of bosonic modes. 
\end{abstract}
\maketitle
\section{Introduction}
Quantum models defined on discrete lattices are very common in solid state theory. Two routes to study their properties are conceivable: Either the attempt to solve the model on the lattice, or the formulation of an effective field theory in the continuum. Lattice models are much more tractable than continuum theories in numerical simulations. On the other hand, within the field-theoretical picture, it is often possible to describe the spectrum in terms of non- or weakly interacting quasiparticles, which makes this approach very attractive from an analytical point of view. Thus, it is most desirable to express the lattice eigenstates in terms of the conceptually much simpler field theoretical eigenstates. 

This goal is generally not achievable, since neither the lattice model nor the full field theory can be solved without approximations. Integrable one-dimensional models, however, are the most promising candidates where such a description can be realized quantitatively. Indeed, {the effective bosonic theory of the spin-1/2 $XXZ$ Heisenberg chain offers the opportunity to obtain the  lattice eigenenergies as an asymptotic expansion in the inverse system length, and at the same time to express the exact lattice eigenstates as linear combinations of bosons}. So far, finite-size corrections to the bosonic spectrum have been determined quantitatively {for particle excited states}.\cite{aff89,luk98} We now calculate the coefficients of the asymptotic corrections quantitatively for {current and particle-hole excited states as well.}

In particular, we obtain the leading terms in an expansion of the lattice energies in the inverse system length to order $\mathcal O\l(N^{-4}\r)$, which translates into a relative deviation of a fraction of a percent or less already for moderate chain lengths $N\sim 20-100$. This is achieved for a large number of low-lying levels, including those for which exact Bethe Ansatz (BA) data are difficult to obtain due to strings of the BA quasimomenta in the complex plane. Furthermore, we express the lattice eigenstates in terms of their bosonic counterparts. Since these states are then eigenstates of a free bosonic theory, our results have the potential to calculate expectation values of local operators in excited states. 

We consider the Hamiltonian of the $XXZ$ model
\be
H=\sum_{j=1}^N(S_j^x S_{j+1}^x+S_j^y S_{j+1}^{y}+\Delta S_j^z S_{j+1}^z),\label{ham}
\ee
with periodic boundary conditions and $N$ lattice sites. We restrict ourselves here to the critical regime $-1<\Delta< 1$; data for the isotropic point $\Delta=1$ are given in the appendix.

By a Jordan-Wigner transformation, Eq.~\refeq{ham} can be mapped to a model for itinerant spinless fermions. The corresponding Hamiltonian reads
\be
H_f&=&\frac12\sum_{j=1}^N\l[c^\dagger_jc_{j+1}+c^\dagger_{j+1}c_{j} + 2\Delta \l(n_j-\frac12\r)\l( n_{j+1}-\frac12\r)\r]\nn\\
& &-\frac{1}{2}\l(1+\te^{\rmi \pi M}\r)\l(c^\dagger_Nc_1+c^\dagger_Nc_1\r)\label{hamf},
\ee 
where $n_j=c^\dagger_j c_j$ and $M$ is the total number of particles, i.e.~the eigenvalue of the total number operator $\sum_{j=1}^N n_j$, which commutes with $H_f$. Thus for an odd (even) number of particles, the boundary conditions of Eq.~\refeq{hamf} are cyclic (anticyclic). The models in Eqs.~\refeq{ham}, \refeq{hamf} have significant experimental
relevance, either in crystals with a strongly aniso\-tropic spin exchange
\cite{ami95,mot96} or in quasi one-dimensional itinerant electron models like
carbon nanotubes.\cite{boc99,ish03,lee04} Most recently, central quantities
like the dynamical structure factor \cite{per06} and the local density
of states \cite{sch07} have been calculated for the lattice model \refeq{ham}
from sums over contributions of individual states.

Historically, the model \refeq{ham} has been studied extensively as a prototypical interacting many body quantum system. The exact solution for $\Delta=-1$ was found by Bethe;\cite{bet31} Hulthen described the isotropic antiferromagnet $\Delta=1$.\cite{hul38}  This solution was generalized to arbitrary $\Delta$ by des Cloizeaux and Gaudin.\cite{clo66} From these works, the ground state and the ground state energy were derived by Yang and Yang.\cite{yan66} Excitations above the ground state were constructed by Takahashi (for a review, see Ref.~[\onlinecite{tak99}]). 

Whereas those works rely on the exact solution of the lattice model, a
field-theoretical approach revealed that excitations with an energy $\Delta E
\ll 1/N$ above the ground state can be described asymptotically, that is in
the limit of large chain lengths, in terms of free quasiparticles that obey bosonic statistics (for a review, see Ref.~[\onlinecite{aff90}]). The corresponding effective Hamiltonian is the Gaussian model, which leads to degeneracies between certain bosonic excitations. Interactions between quasiparticles are captured in additional irrelevant operators,\cite{luk98} which {yield nonlinear contributions to the spectrum} and generally lift the degeneracies. 

Conformal invariance relates the finite-size scaling behavior of each eigenenergy to the scaling dimensions of the operators in
the effective field-theoretical model.\cite{car84a,car84b,car84c} This
connection has been used to predict the scaling dimensions of the leading
irrelevant operators.\cite{alc87,alc88} Here, we employ finite-size scaling to
demonstrate the lifting of degeneracies for individual levels by calculating
the exact contribution of the irrelevant operators to lowest order. 

Therefore, we first concentrate on those low-lying excitation
levels that can be computed numerically from the BA for
arbitrary $N$ without convergence problems. These are mostly parameterized by real BA quasimomenta. After having done
this check, we can use our method to calculate low
excitation energies of the lattice model with an accuracy of $\mathcal
O(N^{-4})$ for $\Delta<1/2$, and $O(N^{-2K})$, $K=\pi/(\pi-\arccos\Delta)$ for
$1/2<\Delta<1$, without using the BA equations, regardless of the underlying BA quasimomenta distribution. Thus especially for
$\Delta<1/2$, we obtain very accurate analytic results for low-lying excitation
energies even of relatively
short chains without having to deal with the BA equations at all,
thus also avoiding string solutions. 

Using this procedure, we then tackle the so far unanswered question of how lattice eigenstates are expressed in terms of bosonic states. This will help to construct the ``physical'' excitations seen in {\em ab initio} numerical methods or experiments as linear combinations of bosons.\cite{sch07}

The remainder of this paper is organized as follows: Section II starts
with a pedagogical introduction into the $\Delta=0$ model and then treats the BA
solution for general $\Delta$. In the third section, the effective low-energy solution from
bosonization is presented, including leading and higher order
contributions. The lattice eigenstates and eigenenergies are expressed
asymptotically through the eigenstates and eigenenergies of the bosonic
low-energy effective Hamiltonian. The relative error in this asymptotic
expansion is illustrated in section IV from a numerical finite size-analysis. 
Appendix A illustrates the conformal towers at the special points $\Delta=0,1/2,1$, and appendix B contains a table that illustrates our labelling
of the BA levels for $N=8$.

\section{The exact lattice solution}
This section summarizes the exact solution of the lattice model. In order to introduce our labeling of the energy levels in the large-$N$-limit, we start with a pedagogical introduction into the non-interacting case $\Delta=0$. The BA solution is presented afterwards. 
\subsection{The free model}
\label{secfree}
\begin{figure}[t]
  \begin{center}
    \includegraphics[width=0.8\textwidth]{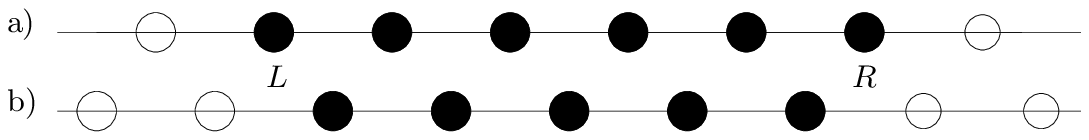}
  \end{center}
  \caption{Sketch of the ground state a) and of a spin excitation $S^z=1$ in b), where one phase is removed and the remaining ones are again grouped symmetrically without vacancies. Here, the ground state a) is composed of an even number of particles, the excited state b) consists of an odd particle number.}
  \label{fig_sp}
\end{figure}

After a Fourier transformation of Eq.~\refeq{hamf}, the energy levels for the free model are given by
\be
E(\l\{k_j\r\})&=&-\sum_{j=1}^{M} \cos k_j , \; S^z=\frac{N}{2}-M\label{enff}\\
k_j&=&\frac{2 \pi n_j}{N}\label{phasmom}
\ee
where the $M$-many phases $n_j$ are chosen out of the {$N$ possible values $-(N+1)/2+p$, $p=1,\ldots,N$}. We denote the values which are not occupied by a phase as ``vacancies''. Here, we restrict ourselves to even $N$, where $S^z=0,1,\ldots,N/2$. The lowest state in each $S^z$-sector (or equivalently, with $M$ many particles) is given by the dense and symmetric distribution of the phases, without any vacancies, see Fig.~\ref{fig_sp}a). This means that the $n_j$ in Eq.~\refeq{phasmom} are integer (half-integer) for $M$ odd (even), corresponding to cyclic (anticyclic) boundary conditions in Eq.~\refeq{hamf}. The ground state energy $E_0$ has no net magnetization, i.e. $M=N/2$, and its leading terms in a large-$N$-expansion are then given by
\be
E_0&=&-\sum_{j=0}^{N/2-1} \cos\frac{\pi}{N} \l(2j-\frac{N}{2}+1\r)=-\frac{1}{\sin\frac\pi
  N}\nn\\
&=&-\frac{N}{\pi}-\frac{\pi}{6N}-\frac{7\pi^3}{360 N^3}+\mathcal
O(N^{-5})\label{gsasymp}\;.
\ee

\begin{figure}[t]
  \begin{center}
    \includegraphics[width=0.8\textwidth]{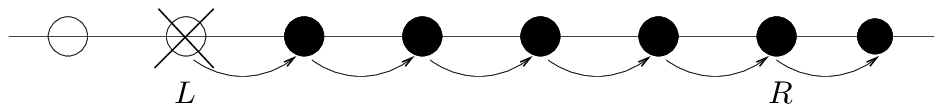}
  \end{center}
  \caption{Sketch of a current excitation $m=1$. The shift of all phases can be either to the right, as depicted, or to the left. The left (right) Fermi points are denoted by $L$ $(R)$, respectively.}
  \label{fig_cur}
\end{figure}
\begin{figure}[t]
  \begin{center}
    \includegraphics[width=0.8\textwidth]{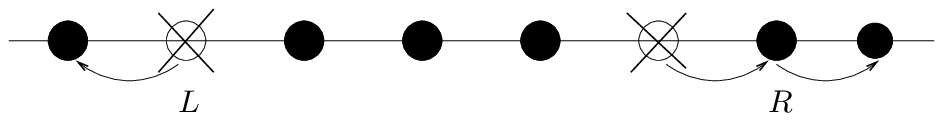}
  \end{center}
  \caption{Sketch of a particle-hole excitation labeled by $(1_1^L, 2_1^R)$. The arrows denote possible shifts in the position of the phases in order to create excitations. The excitations do not have to be symmetric: Allowed are either shifts at the right or the left Fermi point or at both of them. The left (right) Fermi points are denoted by $L$ $(R)$, respectively.}
  \label{fig_bos}
\end{figure}

Since we want to focus on the low-lying excitations above the ground state, it is convenient to introduce a different labeling of the energies. Instead of labeling each energy by the whole set of momenta as in Eq.~\refeq{enff}, we introduce a notation to distinguish between three different types of excitations, according to the distribution of the $n_j$ with respect to the ground state. These labels refer to the true lattice states; their relation to the quantum numbers of the bosonic effective theory will be given in section III.
\begin{itemize}
\item {\em Spin excitations}: The $M$ outermost phases are removed and the remaining ones are again placed symmetrically and without vacancies around the origin, Fig.~\ref{fig_sp}. We will distinguish these excitations by the total spin $S^z=\frac{N}{2}-M$. Due to spin-flip symmetry, the addition of $M$ phases is energetically degenerate to the removal of $M$ phases.
\item {\em Current excitations}: The whole set of numbers is shifted to the right or left by $m$ integers, Fig.~\ref{fig_cur}. We label these excitations by $m$, which we take to be positive (negative) if the shift is to the right (left). 
\item {\em Particle-hole excitations} on top of the ``zero mode'' spin and current excitations: These can always be described by a ``shift'' of occupied states relative to the filled Fermi sea,\cite{egg_ped} labelled by two sets of integers $\{m_n^L\}$, $\{m_n^R\}$, where $m_n^{L,R}$ denotes the number of Fermions that are shifted by $n$ phases at the left ($L$) or right ($R$) Fermi point.  A particular example is given in Fig.~\ref{fig_bos}. These integers resemble bosonic occupation numbers but are so far used as unique labels for lattice eigenstates. The connection to the labels in the continuum bosonic field theory will be established in section \ref{sec2}.
\end{itemize}
In the following, we will label the energy levels by the set of numbers
$(S^z,\,m,\,\{m_n^L\}\cup \{m_n^R\})$. 


In the presence of ``zero mode'' excitations only, the energy of the state with $S^z$-many particles removed from the ground
state and of shifting all the phases by an integer $m$ is obtained as
\be
\!\!&&E(S^z,m,0)-E_0= \l[\cos\l(\frac{\pi S^z}{ N}\r)\,\cos\l(\frac{2m\pi}{N}\r) -1\r] E_0\nn\\
\!\!&&=\frac{\pi}{N}\l[\frac{(S^z)^2}{2}+2 m^2\r]
+\frac{\pi^3}{N^3}\l(\frac{(S^z)^2}{12}-\frac{(S^z)^4}{24}+\frac{m^2}{3}-\frac{2m^4}{3}-(mS^z)^2\r)+\mathcal O(N^{-5})\label{spin_curr_enff}.
\ee
We now look at the particle-hole excitations, first with $S^z=m=0$, and we assume that at the left Fermi point, $N_L$ many phases are involved in the
excitations. Let $j=0$ denote the leftmost phase in the ground state. Then the
pattern of excitations is described by the set $\l\{j_1,\ldots,j_{N_L}\r\}$,
where each integer takes a value $j_i<N_L-1$, including negative integers, and
no two integers are equal. The analogous construction holds for the right
Fermi point. The corresponding energy reads
\be
E_0+\sum_{j=0}^{N_L-1} \sin\frac{2j+1}{N}\pi -
\sum_{\l\{j_1,\ldots,j_{N_L}\r\}} \sin\frac{2j+1}{N} \pi+(L\leftrightarrow R)\label{bos_latt}.
\ee
By considering explicit excitations on a linear spectrum, it can be shown that in leading order in $1/N$ one obtains \cite{egg_ped}
\be
E(0,0,\{m_n^L\}\cup
\{m_n^R\})-E_0=\frac{2\pi}{N} \sum_{n=1}^\infty n(m_n^L+m_n^R)+\mathcal{O}(N^{-3})\label{bosenff}
\ee
Corrections of $\mathcal O(N^{-3})$ cannot be written in terms of $\{m_n^L\}\cup
\{m_n^R\}$ only, as will become clear in section
\ref{sec2}. Most importantly, one sees that the above contributions
\refeq{spin_curr_enff}, \refeq{bosenff} can
be combined linearly in leading order, 
\be
E(S^z,m,\{m_n^L\}\cup \{m_n^R\})-E_0= \frac{2\pi}{N}
\l(\frac{(S^z)^2}{4}+m^2 + \sum_{n=1}^\infty n(m_n^L+m_n^R)\r)+\mathcal{O}(N^{-3}), \label{enff_bos}
\ee
and the higher order terms $\mathcal O(N^{-3})$ contain non-linear contributions where the
excitations mix. These terms will be included on a more general footing in
section \ref{sec2}. 

The particle-hole contribution in Eq.~\refeq{enff_bos} can be considered as the
eigenvalue to the Hamiltonian $\sum_n n \l( a_n^{L\dagger} a_n^L + a_n^{R\dagger}
a_n^R\r)$ with bosonic operators $\l[a_n^{\nu},
  a_m^{\mu\dagger}\r]=\delta_{n,m}\delta_{\nu,\mu}$, where $\nu,\mu$ can be
$R,L$.\cite{aff90,egg_ped} However, due to the higher-order corrections in \refeq{gsasymp}, \refeq{spin_curr_enff}, \refeq{enff_bos}, the labels $(S^z, m,\{m_n^L\}\cup \{m_n^R\})$ introduced in this section are {\em not} the conventional bosonic occupation numbers. Hence the open question arises what the linear combination of the
bosonic eigenstates is that yields the original particle-hole eigenstates of the lattice model.  

In section \ref{sec2} we will show that such mixings of bosonic states can be
determined by taking into account higher-order corrections to \refeq{enff_bos} and
of finite
interactions, $\Delta\neq 0$, if the one or other lead to a splitting of the
corresponding energies. 

\subsection{The Bethe Ansatz solution}
\label{sec1}
In this section, the exact BA solution for the spectrum of the $XXZ$-Hamiltonian with $\Delta\neq 0$ is presented. With its help, we demonstrate how to use the notation for the excitations introduced in the previous section. For original references, we refer the reader to the book by Takahashi [\onlinecite{tak99}].

The energy eigenvalues are parameterized by quasimomenta $k_j$ as
\be
E\l(\l\{k_j\r\}\r)=\frac{\Delta N}{4}-\Delta M - \sum_{j=1}^M \cos k_j\nn
\ee
This equation looks very similar to \refeq{enff}. Now, however, the $M=\frac{N}{2}-S^z$ quasimomenta $k_j$ are solutions to coupled algebraic equations, the Bethe equations
\be
\te^{\rmi k_j N}&=&(-1)^{M-1} \prod_{l\neq j}^M \frac{\exp\l[\rmi(k_j+k_l)\r]+1+2\Delta \exp\l[\rmi k_j\r]}{\exp\l[\rmi(k_j+k_l)\r]+1+2\Delta \exp\l[\rmi k_l\r]}\label{keq}.
\ee
The quasimomenta $k_j$ pertain to interacting magnons above the ferromagnetic state.
For numerical calculations it is more convenient to deal with the logarithmic version of these equations,
\be
k_j N&=&2\pi n_j+2 \sum_{l\neq j}^M \arctan \frac{\Delta \sin\l[(k_j-k_l)/2\r]}{\cos\l[(k_j+k_l)/2\r]+\Delta \cos\l[(k_j-k_l)/2\r]}\label{klog}.
\ee

The lowest energy in the sector with $S^z=N/2-M$ is given by a symmetric choice of the BA numbers in Eq.~(\ref{klog}), such that
\be
n_j=-\frac{M+1}{2}+j, \qquad j=1,\ldots,M\label{phases}
\ee
In the thermodynamic limit, the ground state energy per lattice site, $\ve_0$, is given by
\be
\ve_0&=&\l\{\begin{array}{cc}\frac{\Delta}{4}-\frac{\sin \g}{\g} \i \frac{\sinh\l(\pi/\g -1\r)x}{2\cosh x \,\sinh \pi x/\g} \,\d x,&\;\;\Delta<1\\
\frac14 - \ln 2,&\;\;\Delta=1\end{array}\r.\label{gsen}.
\ee 
We now want to obtain the energies of the lowest excitations directly from the
BA equations. Some of these have been discussed in previous works.\cite{alc87,alc88,aff89,fad81} Our aim here is to perform a systematic study
starting with very small interactions, $|\Delta|\ll 1$, and then to generalize
these results to arbitrary values of $\Delta$. The {distribution of} phases $\{n_j\}$ in Eq.~(\ref{klog}) then defines one state uniquely.

We want to show how to use the labels for the eigenenergies, introduced in the previous section for $\Delta=0$, also in the interacting regime. As a motivation, let us first expand Eq.~\refeq{klog} to first order in $\Delta$, i.e. close to the non-interacting point. This case can still be treated analytically. The corresponding quasimomenta are denoted by $k_j^{(0),(1)}$ and are given by
\be
k_{j}^{(0)}&=&2 \pi \frac{n_j}{N}\label{kj0}\\
k_{j}^{(1)}&=&k_{j}^{(0)} - 2 \frac{\Delta}{N} \sum_l\frac{\sin\l[\l(k_{j}^{(0)}-k_{l}^{(0)}\r)/2\r]}{\cos\l[\l(k_{j}^{(0)}+k_{l}^{(0)}\r)/2\r]}\nn\,.
\ee
This expansion relies on $\Delta \te^{\rmi \l(k_j^{(0)}-k_l^{(0)}\r)/2}/\cos\l[\l(k_j^{(0)}+k_l^{(0)}\r)/2\r]\ll 1$
and therefore has to be taken with care for values $k_j^{(0)}+
k_l^{(0)}\approx \pm\pi$. One special case when this happens is near the Fermi points $k_j^{(0)}\approx
k_l^{(0)}\approx \pm\pi/2$. That the expansion
of the BA equations in the interaction parameter cannot be trusted near the
Fermi points is well known from other BA solvable models.\cite{bat06} However, global quantities which are obtained from summing over
all Bethe numbers, like the energy eigenvalues, turn out to be correct.\cite{bat06, per06} More generally, the condition $k_j^{(0)}+
k_l^{(0)}=\pi$ defines {\em critical pairs},\cite{bie04} with roots that can be
either real or complex. The lowest excited states where these occur are current excitations in the $S^z=0$ sector and particle-hole excitations with $m_n^{L,R}=1_2^{L,R}$, also in the $S^z=0$ sector. A careful analysis shows \cite{bie04} that these
critical pairs can lead to BA numbers different from the phases that one would calculate directly at $\Delta=0$. We illustrate this point in appendix B for a few low-lying states in the chain with $N=8$, $\Delta=1$. However, for any finite $\Delta$, one can still label uniquely each state by the distribution of phases given by $(S^z,m,\{m_n^L\}\cup\{m_n^R\})$ that one would obtain directly at $\Delta=0$, irrespective of the presence of critical pairs. This leads to general expressions for the energy levels to linear order in $\Delta$.  

Namely, the leading terms of the spin excitation energies to linear order in $\Delta$ are given by
\be
E(S^z,0,0)-E_0=\frac{ \pi}{2N}\l(1+\frac{4}{\pi} \Delta\r) \l(S^z\r)^2, \qquad S^z=\frac{N}{2}-M\label{spinen}.
\ee
For the lowest current excitations for $M=N/2$ one obtains the corresponding excitation energy in linear order
\be
E(0,\pm1,0)-E_0=\frac{2\pi}{ N} \label{curex},
\ee
which turns out to be unaffected by $\Delta$ in this order. Finally, the lowest particle-hole excitations have an energy
\be
E(0,0,\{m_n^L\}\cup \{m_n^R\})-E_0=\frac{2 \pi}{N}\l(1-\frac{2}{\pi}\Delta\r)\sum_{n} n(m_n^L + m_n^R)\label{bosen}.
\ee
From Eqs.~\refeq{spinen}-\refeq{bosen} it is clear that $\Delta\neq 0$ generally lifts the degeneracy between the lowest particle-hole and current levels. 


\section{The bosonization solution}
\label{sec2}
In this section, we first review the leading order of the effective bosonic
Hamiltonian for the low-energy excitations which is accurate within $\mathcal
O(N^{-2})$. In the second part, next-leading corrections are included. The aim
of this section is as follows: The eigenstates of
the lattice model, $|S^z,m,\l\{m_n^L\r\}\cup\l\{m_n^R\r\}\rangle_{\rm L}$, are labelled by phase configurations, as described above. On the other hand, as will be made clear below, the eigenstates of the effective model, $|S^z,m,\l\{m_n^L\r\}\cup\l\{m_n^R\r\}\rangle_{\rm B}$, are labelled by the ``zero modes'' and bosonic occupation numbers as derived in [\onlinecite{aff90,egg92}] and shown in Eq.~\refeq{es} below. Here, we wish to
find those linear combinations of the bosonic states that yield the lattice eigenstates.  
\subsection{The leading order: Non-interacting excitations}
Using conventional bosonization, the leading contribution to an effective Hamiltonian, together with its eigenenergies and eigenstates, for the low-energy excitations of
Eq.~\refeq{ham} has been derived.\cite{aff90,egg92} This Gaussian model reads 
\be
\Delta H_0&:=&\lim_{N\to\infty}(H-N \ve_0)\frac{N}{2\pi v}+\frac{1}{12}\\
&=&\frac12\l(\frac{\hat Q^2}{2 \pi} +\frac{\hat \Pi^2}{2 \pi}\r)+\sum_{n=1}^\infty n\l(a_n^{L\dagger}a_n^{L}+a_n^{R\dagger}a_n^{R}\r)\label{hambos}
\ee
with eigenenergies 
\be
\Delta E_0(S^z,m,\{m_n^L\}\cup \{m_n^R\})&=&\lim_{N\to\infty} \l[E(S^z,m,\{m_n^L\}\cup \{m_n^R\})-N \ve_0\r]\frac{N}{2\pi v}+\frac{1}{12}\label{enot}\\
&=&\frac12\l( (S^z)^2/K + K \,m^2 \r)+\sum_{n=1}^\infty n\l(m_n^L+m_n^R\r)\label{ev},
\ee
where $\ve_0$ is the ground state energy per lattice site, given in Eq.~\refeq{gsen}. The effective Hamiltonian and the energies carry an index $_0$ to indicate that they are the leading order in an asymptotic expansion for large $N$ and small $\Delta E\ll 1/N$.

The eigenstates to Eq.~\refeq{hambos} are given by
\be
|S^z,m, \{m_n^L\}\cup \{ m_n^R\}\rangle_{\rm B}&=&\te^{\rmi\l(\sqrt{\frac{2\pi}{K}}S^z \tilde \varphi_0 + \sqrt{2\pi K}m \varphi_0\r)} \prod_{n=1}^\infty \l(a_n^{L\dagger}\r)^{m_n^L}\l(a_n^{R\dagger}\r)^{m_n^R}|0\rangle\label{es},
\ee
with the following commutation relations
\be
&&\l[\varphi_0,\tilde \varphi_0\r]=-\rmi; \;\; \l[\hat Q,\tilde \varphi_0\r]=\rmi; \;\; \l[\hat \Pi,\varphi_0\r]=\rmi;\;\l[a_n^{\mu},a_m^{\nu\dagger}\r]=\delta_{n,m}\delta_{\mu,\nu},\label{commut}
\ee
where $\mu,\nu$ stand for the superscripts $R,L$. 

The exponential in \refeq{es} creates the ``zero mode'' excitations, labelled by the integers $S^z,m$. The product over bosonic operators in \refeq{es} creates bosonic excitations, where the numbers $m_n^{L,R}$ are the bosonic occupation numbers of the $n$-th level. The constants used in Eqs.~(\ref{ev}), (\ref{es}) are
\be
v&=&\frac{\pi}{2}\frac{\sqrt{1-\Delta^2}}{\arccos\Delta}\label{vexp}\\
K&=&\frac{\pi}{\pi-\arccos\Delta}\label{kexp}.
\ee
For weak interactions, $v=1+2 \Delta/\pi + \Or(\Delta^2)$ and $K=2-4\Delta/\pi + \Or(\Delta^2)$, which, together with Eq.~\refeq{ev} agrees with Eqs.~\refeq{spinen}-\refeq{bosen}. 

{In this context, one should note again that in Eqs.~(\ref{enot}), (\ref{ev})  we used the same symbols as in
  Eqs.~(\ref{spin_curr_enff}), (\ref{bosenff}) and
  Eqs.~(\ref{spinen})-(\ref{bosen}) by which - in the asymptotical
  regime - we already identified those energies from the exact solution with
  the ones from bosonization. However, as stated above, the symbols have different meanings: For the lattice eigenstates, they encode the phase configurations, whereas for the bosonic states, they encode bosonic occupation numbers. In the asymptotical regime, the exact
  BA eigenstates are linear combinations of the states \refeq{es} in the
  degenerate subspaces. This will be made explicit in the following section.}

\subsection{Lifting of degeneracies due to irrelevant operators}
The Hamiltonian Eq.~\refeq{hambos} constitutes the leading order in the large $N$-limit. In the bosonization procedure, it results from taking account of spin-density- and spin-current-fluctuations above the ground state, where forward and backward scattering are included. However, Umklapp scattering processes have been neglected so far. Furthermore, Eq.~\refeq{hambos} relies on the linear dispersion approximation of excitations.

Umklapp scattering and non-linear effects in the dispersion relation induce additional terms in the low-energy effective Hamiltonian. These terms are expressed through bosonic fields 
\be
\phi_R(x)&=& \phi_{R,0} + \hat Q_R \frac{x}{\ell} + \sum_{n=1}^\infty \frac{1}{\sqrt{4\pi n}} \l[\te^{2\pi n \rmi x/\ell} a_n^R+\te^{-2\pi n \rmi x/\ell} a_n^{R\dagger}\r]\label{mod1}\\
\phi_L(x)&=& \phi_{L,0} + \hat Q_L \frac{x}{\ell} + \sum_{n=1}^\infty \frac{1}{\sqrt{4\pi n}} \l[\te^{-2\pi n \rmi x/\ell} a_n^L+\te^{2\pi n \rmi x/\ell} a_n^{L\dagger}\r]\label{mod2}\\
\varphi(x)&=&\phi_R(x)+\phi_L(x)\label{mod3}.
\ee
The operators encountered in Eqs.~(\ref{hambos},\ref{commut}) are given by
$\hat Q=\hat Q_R+\hat Q_L$, $\hat\Pi=\hat Q_R-\hat Q_L$, $\varphi_0=\phi_{R,0}+\phi_{L,0}$, $\tilde \varphi_0=\phi_{R,0}-\phi_{L,0}$. 

In the following, leading and nextleading Umklapp processes are encoded in operators $H_c^{(\nu)}$; leading band curvature effects are captured by an operator $H_r$.\cite{luk98,luk03} In our notation, these operators read
\be
H_c^{(\nu)}&=& N\int_{0}^\ell \lambda_\nu \cos\l(\sqrt{8 \pi K}\nu \varphi(x)\r)\d x\label{umkl}\\
H_r&=&N \int_0^\ell \d x\l\{\lambda_+ \l(:\l(\partial_x \phi_R\r)^2:-\frac{\pi}{12
    N^2}\r)\l(:\l(\partial_x \phi_L\r)^2:-\frac{\pi}{12 N^2}\r)\r.\nn\\
& & + \lambda_-\l[:(\partial_x\phi_R)^4: + \frac{3-(1/K+K)}{2\pi}
  :(\partial^2_x \phi_R)^2: + (L\leftrightarrow R)\r. \nn\\
& &\l.\l. +\frac{1}{N^4}\l(\frac{\pi^2}{24}+\frac{\pi^2}{30}(3-(1/K+K))\r)\r]\r\} \label{reg}
\ee
where $\nu$ is a positive integer. Operators in Eq.~\refeq{reg} with $: :$ are normally ordered. For the leading operators
$H_c^{(1)},\,H_r$, the constants $\lambda_1, \,\lambda_+,\,\lambda_-$ are
known,\cite{luk98}
\be
\lambda_1&=& \frac{\Gamma(K) }{2\pi^2 }
\l(\frac{\Gamma(1+1/(2K-2))}{2\sqrt\pi \Gamma(1+K/(2K-2))}\r)^{2K-2}\\
\lambda_-&=&\frac{1}{12\pi K} \frac{\Gamma(3K/(2K-2))
  \Gamma^3(1/(2K-2))}{\Gamma(3/(2K-2))\,\Gamma^3(K/(2K-2))}\\
\lambda_+&=&\frac{1}{2\pi} \,\tan \l[\pi K/(2K-2)\r]\,.
\ee
The constant $\lambda_1$ is given with respect to the CFT-normalization, 
\be
\lim_{N\to\infty} \langle \te^{\rmi \alpha \varphi(x)} \te^{-\rmi \alpha
  \varphi(y)}\rangle=\frac{1}{|x-y|^{2d}}\label{cftnorm}\; ,
\ee
where 
\be
d&=&\frac{\alpha^2}{4\pi}\label{sdim}
\ee
is the scaling dimension of the operator $\te^{\rmi \alpha \varphi}$. The expectation value in Eq.~\refeq{cftnorm} is taken in the ground state.
 
In the following, we will calculate the contribution of the operators \refeq{umkl}, \refeq{reg} in first order perturbation theory. Before going into the details of the calculation, let us first discuss the different contributions that are to be expected from a perturbational treatment. 

\subsubsection{Scaling dimensions and perturbation theory}

The scaling dimension $d$ that governs the behavior of the operators \refeq{hambos}, \refeq{umkl}, \refeq{reg} under RG-transformations can be read off from the exponent of two-point correlation functions like \refeq{cftnorm}. The fixed-point Hamiltonian \refeq{hambos} has scaling dimension $d=2$, whereas the scaling dimension of the leading Umklapp operator \refeq{umkl} is $d=2 K$, see Eq.~\refeq{sdim}. The curvature-like term \refeq{reg} has scaling dimension $d=4$. 

In first order perturbation theory the correction to Eq.~\refeq{ev} from additional operators generally scales like $N^{-(d-2)}$ for finite $N$. Thus the operators \refeq{umkl}, \refeq{reg} can induce additional terms scaling like $N^{-2}$, $N^{-(2K-2)}$, respectively.
\begin{figure}
  \begin{center}
    \includegraphics[width=0.6\textwidth]{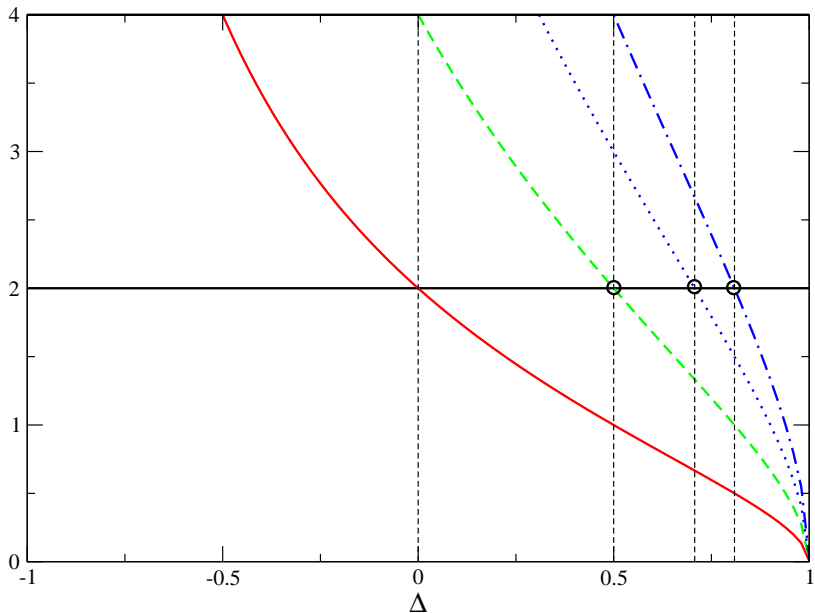}
  \end{center}
  \caption{(Color online) The exponents $2,4$ (horizontal black lines), $2K-2$ (red), $4K-4$
  (green dashed), $6K-6$ and $8K-8$ (blue dotted and dotted-dashed, respectively). The vertical dashed black lines are at the values $\Delta=\cos\frac\pi 2,\,\cos\frac\pi3,\,\cos\frac\pi4,\,\cos\frac\pi5$. The open circles denote the points where logarithmic contributions occur.}
\label{figexp}
\end{figure}

On top of these leading contributions, next-leading terms exist. On the one hand, these stem from second order perturbation theory, giving rise, for example, to terms scaling like $N^{-4}$, $N^{-(4K-4)}$. On the other hand, higher order operators can contribute in first order perturbation theory. For example, the second order Umklapp operator can yield a term $\sim N^{-(8K-2)}$. Generally, any additional operator can create terms in any perturbational order. We illustrate the exponents of the first few leading contributions in Fig.~\ref{figexp}. 

Note that at special values of $\Delta$, the exponents cross. At the free fermion point $\Delta=0$ the amplitudes of the Umklapp operators vanish. However, a non-trivial crossover happens at $\Delta=1/2$. Here, the leading exponent of second order perturbation
theory in $H_c$, $4K-4$, crosses with the exponent 2 stemming from $H_r$. At
this crossover point, the two algebraic corrections merge to form a
logarithmic contribution.\cite{luk98,sir06} The same happens for higher
orders. At the crossover points that are given by roots of unity, $\Delta=\cos\frac{\pi}{n}$, $n> 2$ integer, non-trivial degeneracies between excited levels persist. At these points, the $XXZ$-model has an invariance under the loop algebra $sl_2$, leading to additional degeneracies.\cite{deg01} An unambiguous treatment of these special points has been derived within the BA.\cite{bra01,fab01a,fab01b} In appendix A, we illustrate how these degeneracies show up for the eigenvalues of the Gaussian model, Eq.~\refeq{ev}, by sketching the conformal towers at $\Delta=0,1/2,1$. 
 
We shortly comment on the isotropic case. Obviously, the distinction between leading and next-leading corrections from Umklapp operators does not make sense for the isotropic point, $\Delta=1$. Here, all exponents $2 \nu K -2 \nu\to 0$, and the scaling dimension of $H_c^{(1)}$ is 2 at that point. RG-studies \cite{aff89,nom93,luk98} have shown that {the corrections $\Delta E_1$ to the levels $\Delta E_0$ read}
\be
\Delta E_1(0,0,0)-\Delta E_0(0,0,0)&=& -\frac{3}{8\ln^3 N}\label{ln3},\;\\
\Delta E_1(0,\pm 1,0)-\Delta E_0(0,\pm 1,0)&=& \frac{g_{\pm}}{\ln N},\; \\
\Delta E_1(\pm 1,0,0)-\Delta E_0(\pm 1,0,0)&=& \frac{g_1}{\ln N}.
\ee
The amplitude $g_1$ is known exactly, for the amplitudes $g_{\pm}$, numerical calculations were performed.\cite{aff89} 

In the following, we will concentrate on the first-order contributions of the operators \refeq{umkl}, \refeq{reg}. Especially, we will show that the operator $H_c^{(m)}$ leads to the symmetric/antisymmetric combination of states with $\pm |m|$ and lifts their degeneracy in first order perturbation theory if $S^z=0$. The corrections to Eq.~\refeq{ev} are then of order $N^{2-2 K m^2}$. In all other cases, this operator only contributes in {\em second} order perturbation theory, yielding corrections to Eq.~\refeq{ev} of order $N^{4-4 K m^2}$. The operator \refeq{reg} always contributes in first order perturbation theory, resulting into corrections $\sim N^{-2}$. 

\subsubsection{First-order contributions from Umklapp-operators}
Let us first consider the ``zero mode'' states $|S^z,m,m_n^L=m_n^R=0\rangle_{\rm B, L}$, {constructed according to Eq.~\refeq{es}}. {By inserting the mode expansions \refeq{mod1}-\refeq{mod3} into Eq.~\refeq{umkl} for the leading Umklapp operator $H_c^{(1)}$, one calculates the expectation values of this operator between the ``zero-mode'' excited states. Using the commutation
  relations \refeq{commut} one can show that}
\be
_{\rm B}\langle 0,1,0|H_c^{(1)}|0,1,0\rangle_{\rm B}&=&0\;,\qquad\; _{\rm B}\langle
  0,1,0|H_c^{(1)}|0,-1,0\rangle_{\rm B}=C,\\
_{\rm B}\langle 0,-1,0|H_c^{(1)}|0,-1,0\rangle_{\rm B}&=&0\;,\qquad\; _{\rm B}\langle 0,-1,0|H_c^{(1)}|0,1,0\rangle_{\rm B}=C,  
\ee
with 
\be
C= -(2\pi)^{2 K} \lambda_1 N^{1-2 K}\label{nscal}.
\ee
The corresponding eigenstates in this order are
\be
|0,1,0\rangle_{\rm L}&=&\frac{1}{\sqrt2}\l(|0,1,0\rangle_{\rm
  B}+|0,-1,0\rangle_{\rm B}\r) \nn\\
|0,-1,0\rangle_{\rm L}&=&\frac{1}{\sqrt2}\l(|0,1,0\rangle_{\rm
  B}-|0,-1,0\rangle_{\rm B}\r)\nn\,,
\ee
where the labels of the lattice eigenstates have been chosen according to the
  discussions in section \ref{sec1} and [\onlinecite{bie04}]. Thus
\be
\Delta E_1(0,1,0)&=&-(2\pi)^{2 K} \lambda_1 N^{1-2 K}\label{uken1}\\
\Delta E_1(0,-1,0)&=&(2\pi)^{2 K} \lambda_1 N^{1-2 K}\label{uken2}\; .
\ee
Consequently, the leading operator describing Umklapp scattering lifts the
  degeneracy between the $|m|=1$ states for $S^z=0$ and $m_n^L=m_n^R=0$, such that the
  symmetric combination is energetically lower than the antisymmetric
  combination. 

Within the BA, the interacting quasiparticles above the antiferromagnetic ground state are spinons.\cite{fad81,hal91} Comparing the symmetric and antisymmetric current excitations with the energies of the lowest $S^z=0$
  two-spinon states from the BA,\cite{aff89,bie04} we conclude
  that the symmetric (antisymmetric) combination of current excitations
  corresponds to the lowest two-spinon triplet (singlet) state with $S^z=0$. 

Particle-hole excitations at $|m|=1$ can be included as well. {To determine the expectation values of the Umklapp operator $H_c^{(1)}$ in Eq.~\refeq{umkl} between these states, one again uses Eq.~\refeq{es} together with the commutation relations \refeq{commut}. This results in}
\be
|0,1,1_1^{R,L}\rangle_{\rm L}&=&\frac{1}{\sqrt2}\l(|0,1,1_1^{R,L}\rangle_{\rm
  B}+|0,-1,1_1^{R,L}\rangle_{\rm B}\r)\label{currbos1} \\
|0,-1,1_1^{R,L}\rangle_{\rm L}&=&\frac{1}{\sqrt2}\l((|0,1,1_1^{R,L}\rangle_{\rm
  B}-|0,-1,1_1^{R,L}\rangle_{\rm B}\r)\,,\label{currbos2}
\ee
with energy contributions
\be
\Delta E_1(0,\pm 1,1_1^{R,L})&=& (1-2K)\Delta E_1(0,\pm 1,0)\label{uken3}\\
\Delta E_1(0,\pm 1,1_1^R\,1_1^L)&=& (1-2K)^2\Delta E_1(0,\pm 1,0)\label{uken4}\; .
\ee
The lattice eigenstates are again the symmetric and antisymmetric combinations of the bosonic eigenstates
in this order. 
    
Let us now consider the states $|1,\pm 1,0\rangle_{\rm B}$, that is, states with one spin- and one current-like excitation. {Proceeding similarly as we did in order to arrive at Eqs.~\refeq{uken1},\refeq{uken2}, but now including the additional spin excitation, we find} 
\be
_{\rm B}\langle 1,1,0|H_c^{(1)}|1,1,0\rangle_{\rm B}&=&0\;,\qquad _{\rm B}\langle 1,1,0|H_c^{(1)}|1,-1,0\rangle_{\rm B} =0\;,\label{s11}\\
_{\rm B}\langle 1,-1,0|H_c^{(1)}|1,-1,0\rangle_{\rm B}&=&0\;,\qquad _{\rm B}\langle 1,-1,0|H_c^{(1)}|1,1,0\rangle_{\rm B}=0\label{s12}.    
\ee
Especially, the terms in the second equations in \refeq{s11}, \refeq{s12} now vanish due to the finite magnetization. This argument can be generalized to arbitrary $S^z\neq 0$. Thus we conclude that for states carrying current-like $|m|=1$ {\em and} spin-like excitations, the Umklapp operators $H_c^{(\nu)}$ contribute in {\em second order} perturbation theory only. 

The same is true for states without ``zero mode'', but bosonic excitations only. If these states are degenerate with respect to the Hamiltonian \refeq{hambos}, these degeneracies are not lifted by $H_c^{(1)}$ in first order perturbation theory. However, second order perturbation theory generally yields a finite contribution and can thus lead to a lifting of those degeneracies. 

\subsubsection{First-order contributions from curvature-like terms}
The operator $H_r$ in Eq.~\refeq{reg} yields a finite contribution for all states in first order perturbation
theory. It will generally split the degenerate bosonic levels with the
same excitation energy. In particular, we obtain for the lowest levels by a straightforward evaluation of the expectation values
\be
\Delta E_1(0,0,0)&=&-\frac{\pi^2}{720}\l[5\lambda_++6(5+4b)\lambda_-\r]N^{-2}\label{ev1}\\
\Delta E_1(0,0,1_1^{R,L})&=&\Delta E_1(0,0,0)+\frac{\pi^2}{6}\l[\lambda_++6(1-4b)\lambda_-\r] N^{-2}\label{ev2}\\
\Delta E_1(0,0,1_2^{R,L})&=&\Delta E_1(0,0,0)+\frac{\pi^2}{3}\l[\lambda_+-6(5+4b)\lambda_-
  \r]N^{-2}\label{ev3}\\
\Delta E_1(0,0,2_1^{R,L})&=&\Delta E_1(0,0,0)+\frac{\pi^2}{3}\l[\lambda_++6(1-16b)\lambda_-
  \r]N^{-2}\label{ev4}\\
\Delta E_1(0,0,1_1^R 1_1^L)&=& \Delta E_1(0,0,0)-\frac{\pi^2}{3}\l[11\lambda_+
  -6(1-4b)\lambda_-\r] N^{-2}\label{ev5}\\
\Delta E_1(1,0,0)&=&\frac{\pi^2}{720}\l[
  5\l(\frac1K-6\r)^2\lambda_+\r.\nn\\
& & \l.+\frac{6}{K^2}(60-60 K+(5+4b)K^2)\lambda_-\r]N^{-2}\label{ev6}\\
\Delta E_1(1,0,1_1^{R,L})&=& \Delta E_1(1,0,0)\nn\\
& &-{\pi^2}\l[
  \lambda_+\l(\frac{1}{K}-\frac16\r)+\l(\frac{6}{K}-(1-4b)\r)\lambda_-\r]N^{-2} \label{ev7}\\
\Delta E_1(1,0,1_2^{R,L})&=&\Delta
E_1(1,0,0)+\frac{\pi^2}{3K}\l[(1-6/K) \lambda_+ \r.\nn\\
&&+\l.\l(18\sqrt{8K+(1-2b)^2K^2}-(36+12K+60 b K)\r)\lambda_-\r]N^{-2}\label{ev8}\\
\Delta E_1(1,0,2_1^{R,L})&=&\Delta
E_1(1,0,0)+\frac{\pi^2}{3K}\l[(1-6/K) \lambda_+ \r.\nn\\
&&-\l.\l(18\sqrt{8K+(1-2b)^2K^2}+(36+12K+60 b K)\r)\lambda_-\r]N^{-2}\label{ev9}\\
\Delta E_1(1,0,1_1^{R}1_1^L)&=&\Delta
E_1(1,0,0)\nn\\
& &-\frac{\pi^2}{3K}\l[(6+11K)\lambda_++(36+6(4b-1)K) \lambda_-\r]N^{-2}\label{ev10} .
\ee
Here we defined $b:=3-(1/K+K)$. In this order of $N^{-2}$, with $1/2>\Delta\neq 0$ fixed, the corresponding eigenstates are
\be
|0,0,0\rangle_{\rm L}&=& |0,0,0\rangle_{\rm B}\label{es1}\\
|0,0,1_1^{R,L}\rangle_{\rm L}&=& |0,0,1_1^{R,L}\rangle_{\rm B}\label{es2}\\
|0,0,1_2^{R,L}\rangle_{\rm L}&=& |0,0,1_2^{R,L}\rangle_{\rm B}\label{es3}\\
|0,0,2_1^{R,L}\rangle_{\rm L}&=& |0,0,2_1^{R,L}\rangle_{\rm B}\label{es4}\\
|0,0,1_1^{R}1_1^{L}\rangle_{\rm L}&=& |0,0,1_1^{R}1_1^{L}\rangle_{\rm B}\label{es5}\\
|1,0,0\rangle_{\rm L}&=& |1,0,0\rangle_{\rm B}\label{es6}\\
|1,0,1_1^{R,L}\rangle_{\rm L}&=& |1,1,1_1^{R,L}\rangle_{\rm B}\label{es7}\\
|1,0,1_1^{R}1_1^{L}\rangle_{\rm L}&=&|1,0,1_1^{R}1_1^{L}\rangle_{\rm B}\label{es10}\,.
\ee
Thus for the above low-lying levels, the lattice eigenstates are just the bosonic eigenstates. However, the bosonic states have to be combined appropriately to yield the correct lattice eigenstate for the following levels: 
\be 
|1,0,1_2^{R,L}\rangle_{\rm L}&=& \cos \alpha |0,1,1_2^{R,L}\rangle_{\rm B}+\sin\alpha|0,1,2_1^{R,L}\rangle_{\rm B}\label{es8}\\
|1,0,2_1^{R,L}\rangle_{\rm L}&=& -\sin \alpha |0,1,1_2^{R,L}\rangle_{\rm
  B}+\cos \alpha |0,1,2_1^{R,L}\rangle_{\rm B}\label{es9}\\
\tan \alpha &=&\frac{\sqrt{(2b-1)^2+8/K}+(2b-1)}{\sqrt{(2b-1)^2+8/K}-(2b-1)}\label{angle}.
\ee
In complete analogy, the effect of $H_r$ in Eq.~\refeq{reg} on the current-carrying
states with $|m|=1$ be treated. For example,
\be
E_1(1,\pm1,0)&=&\pm\frac{\pi^2}{720}\l[5\l(2-\frac3K\r)\l(3-\frac2K\r)(6+K(11+6K))\lambda_+\r.\nn\\
& &\l.\frac{6}{K^2}(60-60K+K^2(365+4b+60 K(K-1)))\lambda_-\r]N^{-2}\label{ev11}\\
|1,\pm1,0\rangle_{\rm L}&=& |1,\pm 1,0\rangle_{\rm B}\label{es11}\; .
\ee 

\begin{figure}
\begin{center}
\includegraphics[width=0.6\textwidth]{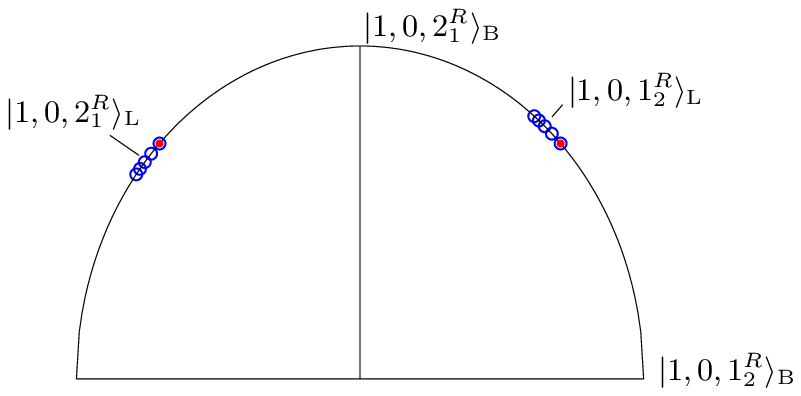} 
\end{center}
\caption{(Color online) Mixing of bosonic states according to (\ref{es8}), (\ref{es9}). The
  circles denote the lattice eigenstates for $\Delta=0,0.1,\ldots,0.4$. The
  red dots are the symmetric and antisymmetric combinations for
  $\Delta=0$.}
\label{rot}
\end{figure}

An important consistency check of \refeq{ev1}-\refeq{ev10} is the limit $\Delta=0$. In this case, $K=2, \, b=1/2, \,\lambda_+=0$ and $\lambda_-=1/6$. Then the above corrections yield those obtained in Eqs.~\refeq{gsasymp}-\refeq{bos_latt}.
(Note, that
the above energy correction has to be multiplied by $2\pi v/N$ to obtain the
contribution to the total energy). The mixing of states in equations
(\ref{es8}), (\ref{es9}) is illustrated in Fig.~\ref{rot}. 

In the BA solution, the energy \refeq{ev3} is encoded by a complex string, which makes a finite-size analysis, especially for large $N$, difficult. The result \refeq{ev3} gives the leading non-universal contribution to this energy analytically, avoiding any problems with strings.  
 
Let us now look at the lattice states $|0,0,1_2^{R,L}\rangle_{\rm L}$, $|0,0,2_1^{R,L}\rangle_{\rm L}$ on the one hand and $|1,0,1_2^{R,L}\rangle_{\rm L}$, $|1,0,2_1^{R,L}\rangle_{\rm L}$ on the other hand. The corresponding lattice and bosonic eigenstates  are given in 
(\ref{es3}), (\ref{es4}) and (\ref{es8}), (\ref{es9}), respectively. For $S^z=1$, the two bosonic states $|1,0,1_2^{R,L}\rangle_{\rm B}$, $|1,0,2_1^{R,L}\rangle_{\rm B}$ are mixed due to $H_r$ with the rotation
angle given in \refeq{angle}. This angle tends to $\pi/4$ in the limit
$\Delta\to 0$, such that (\ref{es8}), (\ref{es9}) are the antisymmetric and
symmetric combinations, respectively. {A similar mixing was found from numerics for the model \refeq{ham} with hard wall boundary conditions.\cite{sch07}} 

{Surprisingly, in this order, the analogous bosonic states for $S^z=0$
  (\ref{es3}), (\ref{es4}) do {\em  not} mix for a given interaction $1/2>\Delta\neq 0$. {The bosonization procedure directly at $\Delta=0$ constructs fermionic states on the lattice that are the symmetric and antisymmetric combinations of the corresponding bosonic states.}\cite{egg_ped} However, at $\Delta=0$, $
  |0,0,1_2^{R,L}\rangle_{\rm L}$ and $|0,0,2_1^{R,L}\rangle_{\rm L}$ are exactly
  degenerate, as can be seen from Eq.~\refeq{bos_latt}, such that any
  combination of them is allowed. A finite
  interaction lifts this degeneracy, but does not mix their bosonic
  equivalents in first order of $H_r$, $H_1$.}

\begin{figure}
  \begin{center}
    \includegraphics[width=0.8\textwidth,angle=0]{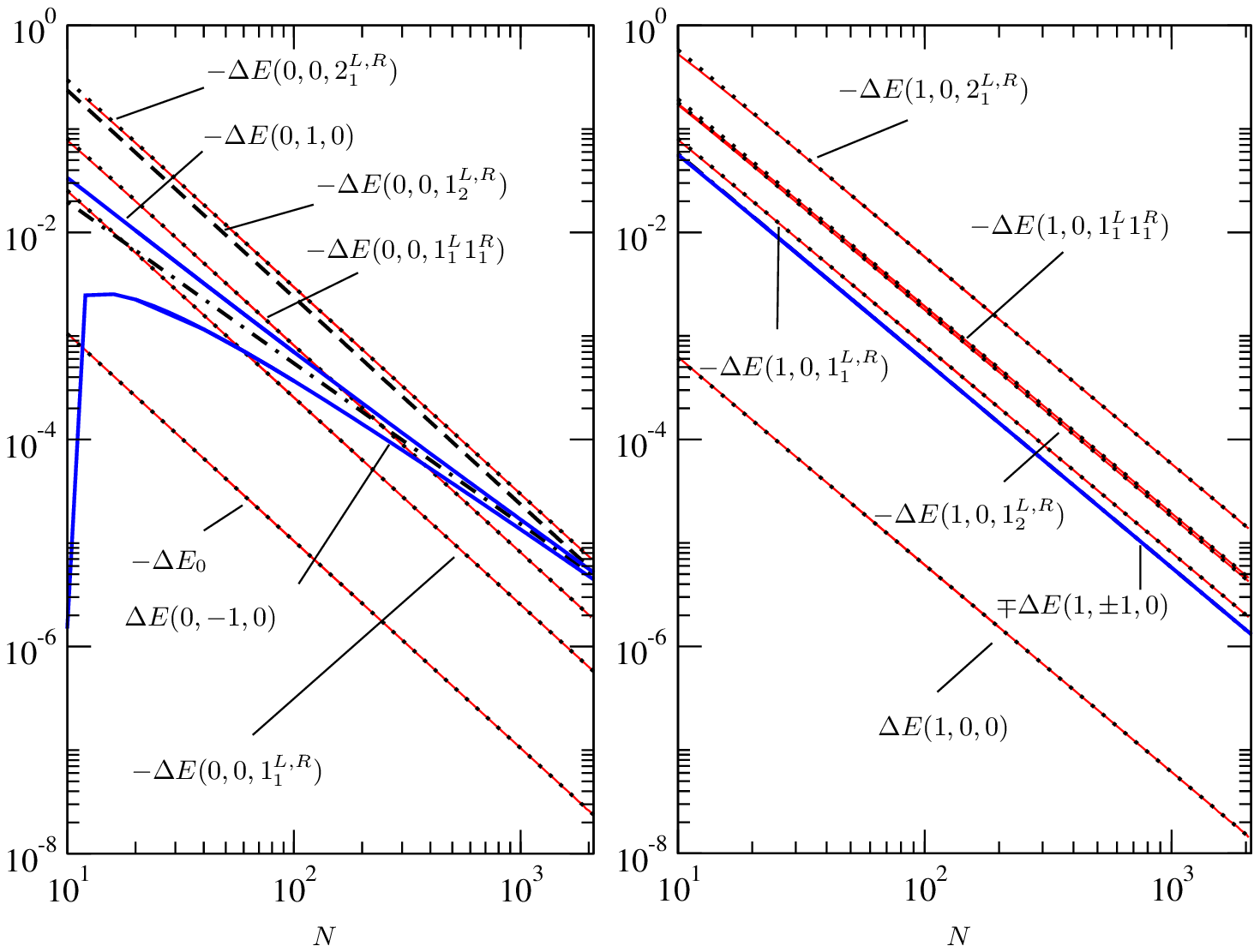}
  \end{center}
  \caption{(Color online) Comparison of the bosonic eigenlevels
  (\ref{ev1})-(\ref{ev10}), dotted black, with the BA values for the corresponding particle-hole excitations, straight red, as a function of $N$ for $\Delta=0.2$. In the left panel, $S^z=0$, and $S^z=1$ in the right panel. The black dashed line in the left panel is the field-theoretical prediction for an energy given by a string-solution in the BA. Also shown (fat blue lines) are current excitations. The dotted-dashed line in the left panel is the field-theoretical result \refeq{uken2}. The field-theoretical result \refeq{ev11} in the right panel cannot be distinguished from the numerical data.}
  \label{fig_num}
\end{figure}

We summarize the above results for the excitation energies and give their scaling behaviors, including the leading perturbational corrections.
\be
\Delta E_1(0,\pm 1, \{m_n^L\}\cup\{m_n^R\})&=&c^{(1)}_c(0,\pm1,m_n^L,m_n^R) N^{2-2
  K}+c_r(0,\pm1,m_n^L,m_n^R)N^{-2}\nn\\
& &+c^{(2)}_c(0,\pm1,m_n^L,m_n^R) N^{4-2 K}\\
\Delta E_1(0,0, \{m_n^L\}\cup\{m_n^R\})&=&c_r(0,0,m_n^L,m_n^R)N^{-2}+c^{(2)}_c(0,0,m_n^L,m_n^R) N^{4-4 K}\\
\Delta E_1(\pm 1,m, \{m_n^L\}\cup\{m_n^R\})&=&c_r(\pm1,m,m_n^L,m_n^R)N^{-2}+c^{(2)}_c(\pm1,m,m_n^L,m_n^R) N^{4-4 K}\nn,\\
\ee
where in the last equation $m=0,\pm 1$. The terms $c_{r}$, $c^{(1)}_c$
are due to first order contributions from $H_r$, $H_1$ and have been
calculated in Eqs.~\refeq{uken1}, \refeq{uken2}, \refeq{uken3}, \refeq{uken4}, \refeq{ev1}-\refeq{ev10}, \refeq{ev11} for the lowest states. Along that way, they can be
determined for any state. The terms $ c^{(2)}_c$
stem from second-order contributions in $H_c$ and are not considered in this
work. For the ground state at given $S^z$, this contribution was determined in Ref.~[\onlinecite{luk98}].

Therefore, the eigenvalues of the Heisenberg chain can be calculated for $\Delta<1/2$
including the order $N^{-{\rm max }(4, 2K)}$ without using the BA
and thus avoiding strings completely. 

\section{Numerical results}

\begin{figure}
  \begin{center}
    \includegraphics[width=0.8\textwidth,angle=0]{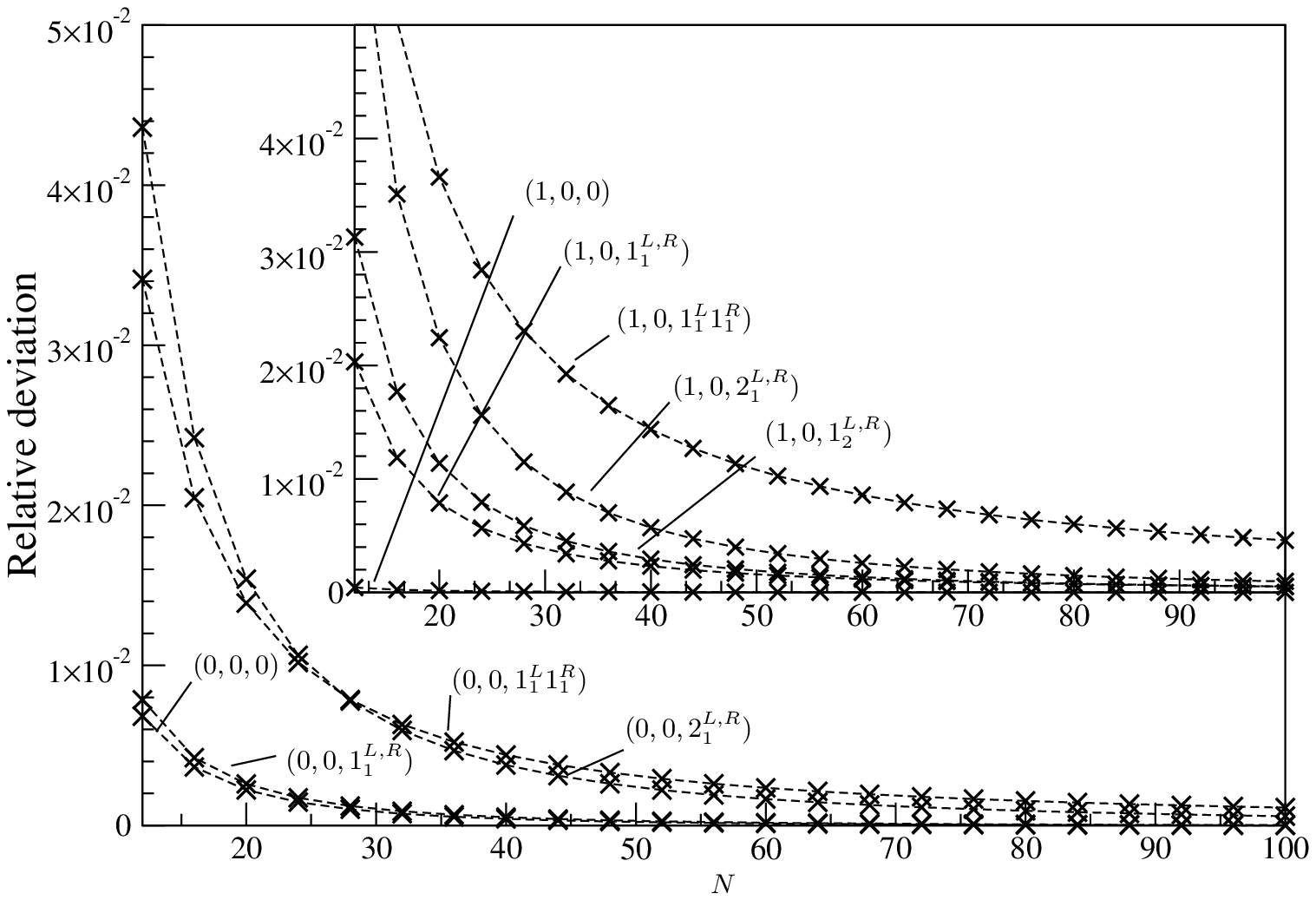}
  \end{center}
  \caption{Relative deviation of the bosonic eigenlevels
  (\ref{ev1}-\ref{ev10}), except for \refeq{ev3}, from the BA values, for
  different ring lengths $N$. The large graph refers to $S^z=0$ levels,
  whereas the inset shows $S^z=1$ levels. The dashed lines are guides to the eye.}
  \label{fig_dev}
\end{figure}
For those states listed in appendix B, we performed a systematic finite-size
scaling analysis up to lattice lengths of $N=2\cdot10^3$ in order to confirm our
results (\ref{uken1})-\refeq{uken4}, \refeq{ev1}-\refeq{ev10} and \refeq{ev11}. 

In Fig.~\ref{fig_num}, we compare the field-theoretical results for a few energies with BA data as a function of the system length $N$ for $\Delta=0.2$. As illustrative examples, we take two current-carrying excitations, Eqs.~\refeq{uken1}, \refeq{uken2} and the bosonic states, \refeq{ev1}-\refeq{ev10}, in the sectors with $S^z=0,1$.

For $S^z=0$, the different exponents of finite-size contributions to the energies of current-carrying excitations compared to states with bosonic excitations only are clearly discernible, see the left panel of Fig.~\ref{fig_num}. On the other hand, for $S^z=1$, the leading finite-size corrections to both current and bosonic excitations scale with the same exponent, as shown in the right panel of Fig.~\ref{fig_num}. 

For all states, the agreement is almost perfect for the longest chains studied. We also give the field-theoretical result for one particle-hole energy which belongs to a complex string in the BA. The field-theoretical prediction is independent of the actual position of roots and thus avoids any convergence problems of complex string solutions in the finite-size scaling analysis. 

In order to estimate the agreement quantitatively, we show the relative deviation in Fig.~\ref{fig_dev}. The plots show
relative deviations of the order of $10^{-2}$ for $N\sim 10$, going down to
roughly $10^{-3}$ for $N\sim 100$. This trend continues, and for $N\sim 2000$,
the relative deviation is around $10^{-4}$ for the energies considered here.  

\section{Conclusion}
We have calculated the coefficients of an asymptotic expansion in the inverse system length of a large number of low-lying excited energies. This calculation does not employ
the BA directly, and thus avoids complex strings which are difficult
to deal with numerically. Instead, the operators \refeq{umkl}, \refeq{reg} have
to be diagonalized which involves the use of bosonic commutation relations
only. This scheme is particularly useful for
$\Delta<1/2$, where the energies are determined analytically within an
accuracy including $\mathcal
O(N^{-{\rm max}(4,2K)})$. For the lowest $\sim 50$ eigenlevels that we have checked, no
degeneracies remain that have not been already present in the lattice model. 

As a further outcome, the lattice eigenstates are expressed in terms of
bosonic modes, again within the accuracy given above, for a fixed
$\Delta$. This representation of eigenstates is different from the BA
representation of eigenstates.  

Applications of this approach can be manifold: Recently, there has been
increasing interest in calculating the dynamic structure factor from a numerical
solution of the BA equations.\cite{per06,kar00,ari06} For other
models, the BA solution is also being used to study dynamical
quantities.\cite{bor07,far08} The calculations presented here to obtain the eigenenergies and
eigenvectors could prove to be useful to obtain both numerical and even
analytical results for form factors restricted to low excitation energies in the
Heisenberg spin chain. The interest in these quantities is high, as underlined by the most recent work,\cite{kit09} where the expectation value of the local magnetization between the ground state and a current-carrying state was computed. 

\section{Acknowledgment}
We thank F.H.L. Essler, S. Reyes and A. Struck for helpful
discussions. Financial support from the Transregional Collaborative Research
Centre SFB/TRR 49 of the {\em Deutsche Forschungsgemeinschaft} and the MATCOR school of excellence is gratefully
acknowledged. M.B. also acknowledges financial support from the European
science network INSTANS and hospitality at the Rudolf-Peierls-Centre for
Theoretical Physics at the University of Oxford, where part of this work has
been carried out.

\appendix

\begin{figure}[b]
  \begin{center}
    \includegraphics[scale=0.65]{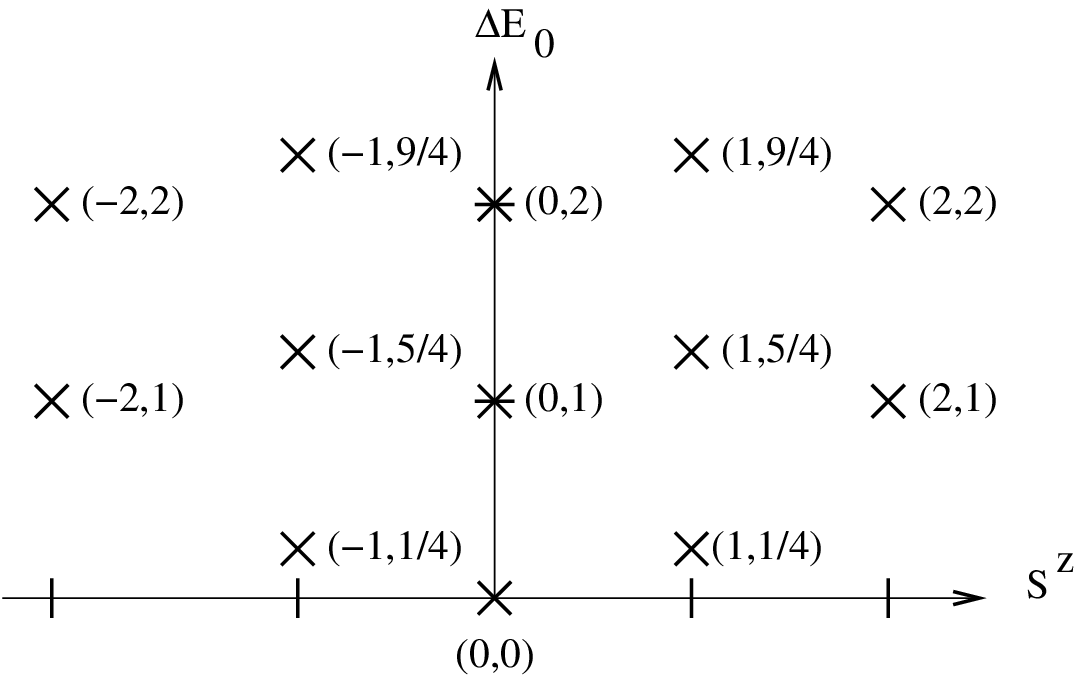}
    \includegraphics[scale=0.65]{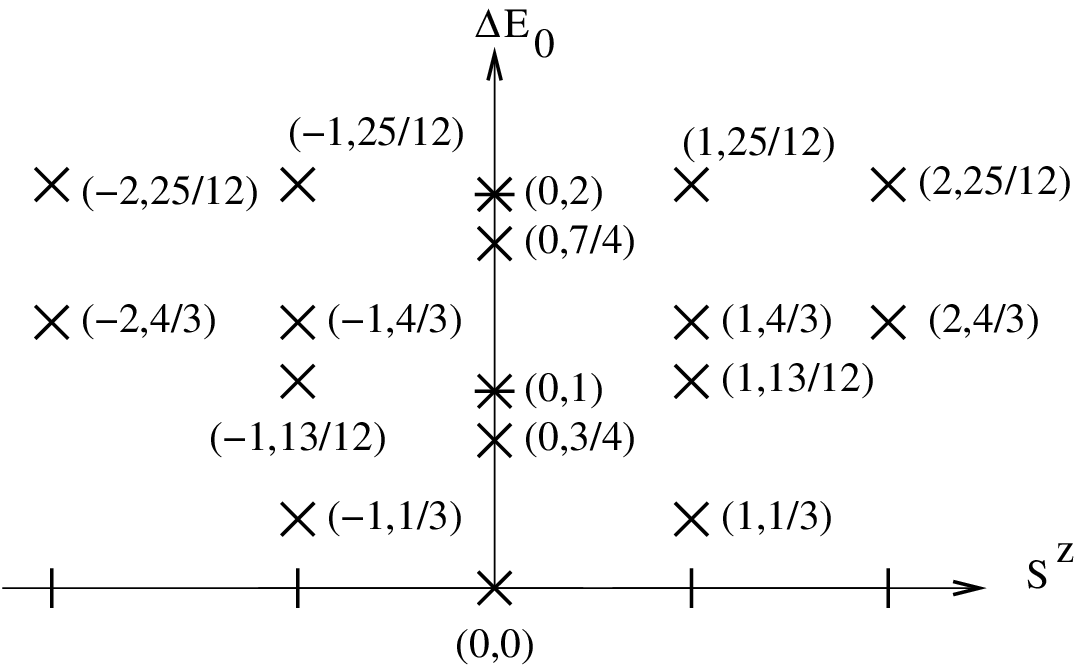}
  \end{center}
  \caption{The lowest excitations for $\Delta=0$ ($\Delta=1/2$) in the left (right) panel. The energies corresponding to the crosses in the $(S^z,\Delta E_0)$-plane are given in \refeq{arr1} and \refeq{arr2}, respectively.} 
  \label{cf_0fig}
\end{figure}
\begin{figure}[h]
  \begin{center}
    \includegraphics[scale=0.65]{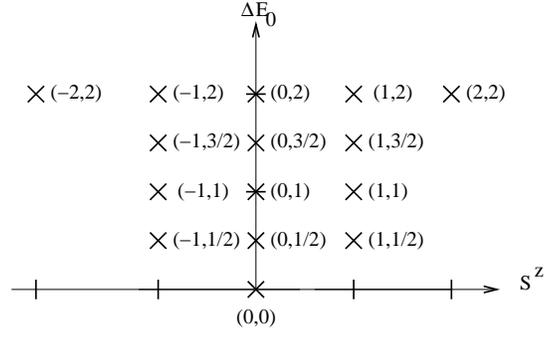}
  \end{center}
  \caption{The lowest excitations for $\Delta=1$. The energies corresponding to the crosses in the $(S^z,\Delta E_0)$-plane are given in \refeq{arr3}. }
  \label{cf_1fig}
\end{figure}
\section{Conformal towers for $\Delta=0,\,1/2,\,1$}
In this appendix, we illustrate the low-energy spectra \refeq{ev} for $\Delta=0,\,1/2,\,1$ in an $S^z-\Delta E_0$-diagram (conformal tower). On the one hand, this shows the lifting of degeneracies at finite $\Delta$ compared to the $\Delta=0$-case. On the other hand, it demonstrates the occurrence of non-trivial symmetries at the special points $\Delta=1/2=\cos\frac{\pi}{3}$ and $\Delta=1$. 

Each cross in a conformal tower is labeled by its coordinates in the $(S^z,\,\Delta E_0)$ plane. The corresponding quantum numbers $(S^z, m,\{m_n^L\}\cup \{m_n^R\})$ of the bosonic field theory are listed in the following. 
\begin{itemize}
\item For $\Delta=0$: \be\begin{array}{ll}
(0,1):& (0,0,1_1^L), (0,0,1_1^R), (0,\pm 1,0);\\
 (0,2):& (0,0,1_1^L \,1_1^R), (0,0,2_1^L), (0,0,2_1^R), (0,0,1_2^R), (0,0,1_2^L), (0,\pm 1,1_1^L), (0,\pm 1,1_1^R);\\
 (\pm 1,1/4):& (\pm 1,0,0); \\
 (\pm 1,5/4):& (\pm 1,0,1_1^L), (\pm 1,0,1_1^R), (\pm 1,\pm 1,0);\\
 (\pm 1,9/4):& (\pm 1,0,1_1^L\,1_1^R), (\pm 1,0,2_1^L), (\pm 1,0,2_1^R), (\pm 1,0,1_2^L), (\pm 1,0,1_2^R), \\
&(\pm 1,\pm 1,1_1^L), (\pm 1,\pm 1,1_1^R);\\
 (\pm 2,1):& (\pm 2,0,0); \\
(\pm 2,2):& (\pm 2,0,1_1^L), (\pm 2,0,1_1^R), (\pm 2,\pm 1,0)\end{array}\label{arr1}
\ee
\item For $\Delta=1/2$: 
  \be\begin{array}{ll}
(0,3/4):& (0,\pm 1,0);\\
 (0,1):& (0,0,1_1^L),(0,0,1_1^R);\\
 (0,7/4) :& (0,\pm 1,1_1^L), (0,\pm 1,1_1^R);\\
 (0,2):& (0,0,1_1^L\,1_1^R), (0,0,2_1^L), (0,0,2_1^R), (0,0,1_2^L), (0,0,1_2^R);\\
 (\pm 1,1/3):& (\pm 1,0,0);\\
 (\pm 1,13/12):& (\pm 1,\pm 1,0); \\
(\pm 1,4/3) :& (\pm 1,0,1_1^L), (\pm 1,0,1_1^R);\\
 (\pm 1,25/12):& (\pm 1,\pm 1,1_1^L), (\pm 1,\pm 1,1_1^R);\\
 (\pm 2,4/3):&(\pm 2,0,0);\\
 (\pm 2, 25/12):& (\pm 2,\pm 1,0)\end{array}\label{arr2}\ee
\item For $\Delta=1$: 
\be
\begin{array}{ll}
(0.1/2):& (0,\pm 1,0);\\
 (0,1):& (0,0,1_1^L),(0,0,1_1^R);\\
 (0,3/2) :& (0,\pm 1,1_1^L), (0,\pm 1,1_1^R);\\
 (0,2):& (0,0,1_1^L\,1_1^R), (0,0,2_1^L), (0,0,2_1^R), (0,0,1_2^L), (0,0,1_2^R), (0,\pm 2,0);\\
 (\pm 1,1/2):& (\pm 1,0,0);\\
 (\pm 1,1):& (\pm 1,\pm 1,0);\\
 (\pm 1,3/2) :& (\pm 1,0,1_1^L), (\pm 1,0,1_1^R);\\
 (\pm 1,2):& (\pm 1,\pm 1,1_1^L), (\pm 1,\pm 1,1_1^R);\\
 (\pm 2,2):&(\pm 2,0,0)\end{array}\label{arr3}\ee
\end{itemize}

\section{Low energy states for the $XXX$-chain with $N=8$ lattice sites} 
\label{appb}
In the following table, a few low-lying excited states above the ground state (note the
$\pm S^z$ symmetry) are given
for the $XXX$-chain with 8 lattice sites. The BA numbers are given as well as
the quasimomenta and the lattice labels, according to section \ref{sec1}. {For excitations with purely real quasimomenta, that is for phases $|n_j|<n_c$ with $n_c=(S^z+N/2)/2$, there are no convergence problems in the finite size scaling analysis because all quasimomenta are real.\cite{bie04} Otherwise, string solutions occur, which have to be treated separately. The
stars $*$ symbolize the occurrence of such solutions, where a critical pair forms either a real
or a complex string.} These are separated off from the BA equations. Ref.~[\onlinecite{bie04}] shows how to assign BA numbers to these critical pairs as well.  
{\small
\be
\begin{array}{l|l|rrrr|llll|l}
E& S^z& \multicolumn{4}{c}{2 n_j}\vline & \multicolumn{4}{c}{k_j}\vline &
\mbox{Lattice label}\\
\hline
-3.65109&0&-3 & -1 & 1 & 3 & -1.61959&-0.506761&0.506761&1.61959& (0,0,0)\\
-3.12842&0& -1  &  1    & *    & *    & 0&\pi&-0.971321& 0.971321&(0,1,0)\\
-3.12842&1& -2  &  0   &  2  &     & -0.971321&0&-0.971321 &&(1,0,0)\\
-2.69963&0&  -1   & 1     &   *  & *    &\frac\pi2+ \rmi\infty&\frac\pi2-\rmi\infty&-0.555164& 0.555164&(0,-1,0)\\
-2.45874&0&\mp3 &\mp1  &\pm1  &\pm5  &\pm\pi&\pm0.214056&\pm0.738788&\pm1.83146&(0,0,1_1^{R,L})\\
-2.45874&1&0 & \pm2 & \pm4 & & \mp0.214056& \pm0.738788& \pm1.83146& & (1,\pm 1,0)\\
-2.14515&1&\mp2&0&\pm4& & \mp1.04767& \mp0.0951988& \pm1.92827& & (1,0,1_1^{R,L})\\
-2.14515&0&\mp 5 &\mp 3&\mp 1&\pm 3& \mp\pi&\mp1.04767& \mp0.0951988& \pm1.92827&(0,\mp 1, 1_1^{R,L}) \\
-1.85464&0&\mp3&\mp1&\pm3&\pm5&\mp1.87588& \mp0.794255& \pm1.09933&\pm\pi&(0,0,2_1^{R,L})\\
-1.85454&1&\mp2&\pm2&\pm4& & \mp 1.09933& \pm0.794255& \pm1.87588& & (1,\mp1,1_1^{R,L})\\
-1.80194&2&-1 & 1 & & & -0.448799& 0.448799& & &(2,0,0)\\
-1.80194&0&-5&-1&1&5&-\pi&-0.448799& 0.448799&\pi &(0,0,1_1^L1_1^R)\\
-1.80194&1 &0  &\pm 2  &\pm  6 & &\mp 0.448799    &\pm  0.448799   & \pm \pi        &         &   (1,\pm 1,1_1^{R,L})             \\
\end{array}\nn
\ee}


\begin{thebibliography}{39}
\expandafter\ifx\csname natexlab\endcsname\relax\def\natexlab#1{#1}\fi
\expandafter\ifx\csname bibnamefont\endcsname\relax
  \def\bibnamefont#1{#1}\fi
\expandafter\ifx\csname bibfnamefont\endcsname\relax
  \def\bibfnamefont#1{#1}\fi
\expandafter\ifx\csname citenamefont\endcsname\relax
  \def\citenamefont#1{#1}\fi
\expandafter\ifx\csname url\endcsname\relax
  \def\url#1{\texttt{#1}}\fi
\expandafter\ifx\csname urlprefix\endcsname\relax\def\urlprefix{URL }\fi
\providecommand{\bibinfo}[2]{#2}
\providecommand{\eprint}[2][]{\url{#2}}

\bibitem[{\citenamefont{Lukyanov}(1998)}]{luk98}
\bibinfo{author}{\bibfnamefont{S.}~\bibnamefont{Lukyanov}},
  \bibinfo{journal}{Nucl. Phys. B} \textbf{\bibinfo{volume}{522}},
  \bibinfo{pages}{533} (\bibinfo{year}{1998}).

\bibitem[{\citenamefont{Affleck et~al.}(1989)\citenamefont{Affleck, Gepner,
  Schulz, and Ziman}}]{aff89}
\bibinfo{author}{\bibfnamefont{I.}~\bibnamefont{Affleck}},
  \bibinfo{author}{\bibfnamefont{D.}~\bibnamefont{Gepner}},
  \bibinfo{author}{\bibfnamefont{H.}~\bibnamefont{Schulz}}, \bibnamefont{and}
  \bibinfo{author}{\bibfnamefont{T.}~\bibnamefont{Ziman}}, \bibinfo{journal}{J.
  Phys. A} \textbf{\bibinfo{volume}{22}}, \bibinfo{pages}{551}
  (\bibinfo{year}{1989}).

\bibitem[{\citenamefont{Ami et~al.}(1995)\citenamefont{Ami, K., L., Wang,
  Johnston, Huang, and Erwin}}]{ami95}
\bibinfo{author}{\bibfnamefont{T.}~\bibnamefont{Ami}},
  \bibinfo{author}{\bibfnamefont{C.~M.} \bibnamefont{K.}},
  \bibinfo{author}{\bibfnamefont{H.~R.} \bibnamefont{L.}},
  \bibinfo{author}{\bibfnamefont{Z.~R.} \bibnamefont{Wang}},
  \bibinfo{author}{\bibfnamefont{D.~C.} \bibnamefont{Johnston}},
  \bibinfo{author}{\bibfnamefont{Q.}~\bibnamefont{Huang}}, \bibnamefont{and}
  \bibinfo{author}{\bibfnamefont{R.~D.} \bibnamefont{Erwin}},
  \bibinfo{journal}{Phys. Rev. B} \textbf{\bibinfo{volume}{51}},
  \bibinfo{pages}{5994} (\bibinfo{year}{1995}).

\bibitem[{\citenamefont{Motoyama et~al.}(1996)\citenamefont{Motoyama, Eisaki,
  and Uchida}}]{mot96}
\bibinfo{author}{\bibfnamefont{N.}~\bibnamefont{Motoyama}},
  \bibinfo{author}{\bibfnamefont{H.}~\bibnamefont{Eisaki}}, \bibnamefont{and}
  \bibinfo{author}{\bibfnamefont{S.}~\bibnamefont{Uchida}},
  \bibinfo{journal}{Phys. Rev. Lett.} \textbf{\bibinfo{volume}{76}},
  \bibinfo{pages}{3212} (\bibinfo{year}{1996}).

\bibitem[{\citenamefont{Bockrath et~al.}(1999)\citenamefont{Bockrath, Cobden,
  Lu, Rinzler, Smalley, Balents, and McEuen}}]{boc99}
\bibinfo{author}{\bibfnamefont{M.}~\bibnamefont{Bockrath}},
  \bibinfo{author}{\bibfnamefont{D.~H.} \bibnamefont{Cobden}},
  \bibinfo{author}{\bibfnamefont{J.}~\bibnamefont{Lu}},
  \bibinfo{author}{\bibfnamefont{A.}~\bibnamefont{Rinzler}},
  \bibinfo{author}{\bibfnamefont{R.~E.} \bibnamefont{Smalley}},
  \bibinfo{author}{\bibfnamefont{L.}~\bibnamefont{Balents}}, \bibnamefont{and}
  \bibinfo{author}{\bibfnamefont{P.~L.} \bibnamefont{McEuen}},
  \bibinfo{journal}{Nature} \textbf{\bibinfo{volume}{397}},
  \bibinfo{pages}{598} (\bibinfo{year}{1999}).

\bibitem[{\citenamefont{Ishii et~al.}(2003)\citenamefont{Ishii, Kataura,
  Shiozawa, Yoshioka, Otsubo, Takayama, Miyahara, Suzuki, Achiba, Nakatake
  et~al.}}]{ish03}
\bibinfo{author}{\bibfnamefont{H.}~\bibnamefont{Ishii}},
  \bibinfo{author}{\bibfnamefont{H.}~\bibnamefont{Kataura}},
  \bibinfo{author}{\bibfnamefont{H.}~\bibnamefont{Shiozawa}},
  \bibinfo{author}{\bibfnamefont{H.}~\bibnamefont{Yoshioka}},
  \bibinfo{author}{\bibfnamefont{H.}~\bibnamefont{Otsubo}},
  \bibinfo{author}{\bibfnamefont{Y.}~\bibnamefont{Takayama}},
  \bibinfo{author}{\bibfnamefont{T.}~\bibnamefont{Miyahara}},
  \bibinfo{author}{\bibfnamefont{S.}~\bibnamefont{Suzuki}},
  \bibinfo{author}{\bibfnamefont{Y.}~\bibnamefont{Achiba}},
  \bibinfo{author}{\bibfnamefont{M.}~\bibnamefont{Nakatake}},
  \bibnamefont{et~al.}, \bibinfo{journal}{Nature}
  \textbf{\bibinfo{volume}{426}}, \bibinfo{pages}{540} (\bibinfo{year}{2003}).

\bibitem[{\citenamefont{Lee et~al.}(2004)\citenamefont{Lee, Eggert, Kim, Kahng,
  Shinoara, and Kuk}}]{lee04}
\bibinfo{author}{\bibfnamefont{J.}~\bibnamefont{Lee}},
  \bibinfo{author}{\bibfnamefont{S.}~\bibnamefont{Eggert}},
  \bibinfo{author}{\bibfnamefont{H.}~\bibnamefont{Kim}},
  \bibinfo{author}{\bibfnamefont{S.-J.} \bibnamefont{Kahng}},
  \bibinfo{author}{\bibfnamefont{H.}~\bibnamefont{Shinoara}}, \bibnamefont{and}
  \bibinfo{author}{\bibfnamefont{Y.}~\bibnamefont{Kuk}},
  \bibinfo{journal}{Phys. Rev. Lett.} \textbf{\bibinfo{volume}{93}},
  \bibinfo{pages}{166403} (\bibinfo{year}{2004}).

\bibitem[{\citenamefont{Pereira et~al.}(2006)\citenamefont{Pereira, Sirker,
  Caux, Hagemans, Maillet, White, and Affleck}}]{per06}
\bibinfo{author}{\bibfnamefont{R.}~\bibnamefont{Pereira}},
  \bibinfo{author}{\bibfnamefont{J.}~\bibnamefont{Sirker}},
  \bibinfo{author}{\bibfnamefont{J.-S.} \bibnamefont{Caux}},
  \bibinfo{author}{\bibfnamefont{R.}~\bibnamefont{Hagemans}},
  \bibinfo{author}{\bibfnamefont{J.}~\bibnamefont{Maillet}},
  \bibinfo{author}{\bibfnamefont{S.}~\bibnamefont{White}}, \bibnamefont{and}
  \bibinfo{author}{\bibfnamefont{I.}~\bibnamefont{Affleck}},
  \bibinfo{journal}{Phys. Rev. Lett.} \textbf{\bibinfo{volume}{96}},
  \bibinfo{pages}{257202} (\bibinfo{year}{2006});
  \bibinfo{journal}{J. Stat. Mech.} p. \bibinfo{pages}{P08022}
  (\bibinfo{year}{2007}).

\bibitem[{\citenamefont{Schneider et~al.}(2008)\citenamefont{Schneider, Bortz,
  Struck, and Eggert}}]{sch07}
\bibinfo{author}{\bibfnamefont{I.}~\bibnamefont{Schneider}},
  \bibinfo{author}{\bibfnamefont{M.}~\bibnamefont{Bortz}},
  \bibinfo{author}{\bibfnamefont{A.}~\bibnamefont{Struck}}, \bibnamefont{and}
  \bibinfo{author}{\bibfnamefont{S.}~\bibnamefont{Eggert}},
  \bibinfo{journal}{Phys. Rev. Lett.} \textbf{\bibinfo{volume}{101}},
  \bibinfo{pages}{206401} (\bibinfo{year}{2008}).

\bibitem[{\citenamefont{Bethe}(1931)}]{bet31}
\bibinfo{author}{\bibfnamefont{H.}~\bibnamefont{Bethe}}, \bibinfo{journal}{Z.
  Phys.} \textbf{\bibinfo{volume}{71}}, \bibinfo{pages}{205}
  (\bibinfo{year}{1931}).

\bibitem[{\citenamefont{Hulthen}(1938)}]{hul38}
\bibinfo{author}{\bibfnamefont{L.}~\bibnamefont{Hulthen}},
  \bibinfo{journal}{Arkiv Math. Astron. Fys.} \textbf{\bibinfo{volume}{26A}}
  (\bibinfo{year}{1938}).

\bibitem[{\citenamefont{des Cloizeaux and Gaudin}(1966)}]{clo66}
\bibinfo{author}{\bibfnamefont{J.}~\bibnamefont{des Cloizeaux}}
  \bibnamefont{and} \bibinfo{author}{\bibfnamefont{M.}~\bibnamefont{Gaudin}},
  \bibinfo{journal}{J. Math. Phys.} \textbf{\bibinfo{volume}{7}},
  \bibinfo{pages}{1384} (\bibinfo{year}{1966}).

\bibitem[{\citenamefont{Yang and Yang}(1966)}]{yan66}
\bibinfo{author}{\bibfnamefont{C.}~\bibnamefont{Yang}} \bibnamefont{and}
  \bibinfo{author}{\bibfnamefont{C.}~\bibnamefont{Yang}},
  \bibinfo{journal}{Phys. Rev.} \textbf{\bibinfo{volume}{150}},
  \bibinfo{pages}{321} (\bibinfo{year}{1966}).

\bibitem[{\citenamefont{Takahashi}(1999)}]{tak99}
\bibinfo{author}{\bibfnamefont{M.}~\bibnamefont{Takahashi}},
  \emph{\bibinfo{title}{Thermodynamics of one-dimensional solvable problems}}
  (\bibinfo{publisher}{Cambridge University Press}, \bibinfo{year}{1999}).

\bibitem[{\citenamefont{Affleck}(1990)}]{aff90}
\bibinfo{author}{\bibfnamefont{I.}~\bibnamefont{Affleck}}, in
  \emph{\bibinfo{booktitle}{Fields, Strings and Critical Phenomena}}, edited by
  \bibinfo{editor}{\bibfnamefont{E.}~\bibnamefont{Br\'ezin}} \bibnamefont{and}
  \bibinfo{editor}{\bibfnamefont{J.}~\bibnamefont{Zinn-Justin}}
  (\bibinfo{publisher}{North Holland, Amsterdam}, \bibinfo{year}{1990}), p.
  \bibinfo{pages}{563}.

\bibitem[{\citenamefont{Cardy}(1984{\natexlab{a}})}]{car84a}
\bibinfo{author}{\bibfnamefont{J.~L.} \bibnamefont{Cardy}},
  \bibinfo{journal}{J. Phys. A} \textbf{\bibinfo{volume}{17}},
  \bibinfo{pages}{L385} (\bibinfo{year}{1984}{\natexlab{a}}).

\bibitem[{\citenamefont{Cardy}(1984{\natexlab{b}})}]{car84b}
\bibinfo{author}{\bibfnamefont{J.~L.} \bibnamefont{Cardy}},
  \bibinfo{journal}{J. Phys. A} \textbf{\bibinfo{volume}{17}},
  \bibinfo{pages}{L957} (\bibinfo{year}{1984}{\natexlab{b}}).

\bibitem[{\citenamefont{Cardy}(1984{\natexlab{c}})}]{car84c}
\bibinfo{author}{\bibfnamefont{J.~L.} \bibnamefont{Cardy}},
  \bibinfo{journal}{Nucl. Phys. B} \textbf{\bibinfo{volume}{240}},
  \bibinfo{pages}{514} (\bibinfo{year}{1984}{\natexlab{c}}).

\bibitem[{\citenamefont{Alcaraz et~al.}(1987)\citenamefont{Alcaraz, Barber, and
  Batchelor}}]{alc87}
\bibinfo{author}{\bibfnamefont{F.~C.} \bibnamefont{Alcaraz}},
  \bibinfo{author}{\bibfnamefont{M.}~\bibnamefont{Barber}}, \bibnamefont{and}
  \bibinfo{author}{\bibfnamefont{M.}~\bibnamefont{Batchelor}},
  \bibinfo{journal}{Phys. Rev. Lett.} \textbf{\bibinfo{volume}{58}},
  \bibinfo{pages}{771} (\bibinfo{year}{1987}).

\bibitem[{\citenamefont{Alcaraz et~al.}(1988)\citenamefont{Alcaraz, Barber, and
  Batchelor}}]{alc88}
\bibinfo{author}{\bibfnamefont{F.~C.} \bibnamefont{Alcaraz}},
  \bibinfo{author}{\bibfnamefont{M.}~\bibnamefont{Barber}}, \bibnamefont{and}
  \bibinfo{author}{\bibfnamefont{M.}~\bibnamefont{Batchelor}},
  \bibinfo{journal}{Ann. Phys.} \textbf{\bibinfo{volume}{182}},
  \bibinfo{pages}{280} (\bibinfo{year}{1988}).

\bibitem[{\citenamefont{Eggert}(2007)}]{egg_ped}
\bibinfo{author}{\bibfnamefont{S.}~\bibnamefont{Eggert}}, in
  \emph{\bibinfo{booktitle}{Theoretical Survey of One Dimensional Wire
  Systems}}, edited by \bibinfo{editor}{\bibfnamefont{Y.}~\bibnamefont{Kuk}}
  \bibnamefont{and} \bibinfo{editor}{\bibnamefont{et~al.}}
  (\bibinfo{publisher}{Sowha Publishing, Seoul}, \bibinfo{year}{2007}).

\bibitem[{\citenamefont{Faddeev and Takhtajan}(1981)}]{fad81}
\bibinfo{author}{\bibfnamefont{L.}~\bibnamefont{Faddeev}} \bibnamefont{and}
  \bibinfo{author}{\bibfnamefont{L.}~\bibnamefont{Takhtajan}},
  \bibinfo{journal}{Phys. Lett. A} \textbf{\bibinfo{volume}{85}},
  \bibinfo{pages}{375} (\bibinfo{year}{1981}).

\bibitem[{\citenamefont{Batchelor et~al.}(2006)\citenamefont{Batchelor, Bortz,
  Oelkers, and Guan}}]{bat06}
\bibinfo{author}{\bibfnamefont{M.}~\bibnamefont{Batchelor}},
  \bibinfo{author}{\bibfnamefont{M.}~\bibnamefont{Bortz}},
  \bibinfo{author}{\bibfnamefont{N.}~\bibnamefont{Oelkers}}, \bibnamefont{and}
  \bibinfo{author}{\bibfnamefont{X.-W.} \bibnamefont{Guan}},
  \bibinfo{journal}{J. Phys.: Conf. Ser.} \textbf{\bibinfo{volume}{42}},
  \bibinfo{pages}{5} (\bibinfo{year}{2006}).

\bibitem[{\citenamefont{Biegel et~al.}(2004)\citenamefont{Biegel, Karbach,
  M\"uller, and Wiele}}]{bie04}
\bibinfo{author}{\bibfnamefont{D.}~\bibnamefont{Biegel}},
  \bibinfo{author}{\bibfnamefont{M.}~\bibnamefont{Karbach}},
  \bibinfo{author}{\bibfnamefont{G.}~\bibnamefont{M\"uller}}, \bibnamefont{and}
  \bibinfo{author}{\bibfnamefont{K.}~\bibnamefont{Wiele}},
  \bibinfo{journal}{Phys. Rev. B} \textbf{\bibinfo{volume}{69}},
  \bibinfo{pages}{174404} (\bibinfo{year}{2004}).

\bibitem[{\citenamefont{Eggert and Affleck}(1992)}]{egg92}
\bibinfo{author}{\bibfnamefont{S.}~\bibnamefont{Eggert}} \bibnamefont{and}
  \bibinfo{author}{\bibfnamefont{I.}~\bibnamefont{Affleck}},
  \bibinfo{journal}{Phys. Rev. B} \textbf{\bibinfo{volume}{46}},
  \bibinfo{pages}{10866} (\bibinfo{year}{1992}).

\bibitem[{\citenamefont{Lukyanov and Terras}(2003)}]{luk03}
\bibinfo{author}{\bibfnamefont{S.}~\bibnamefont{Lukyanov}} \bibnamefont{and}
  \bibinfo{author}{\bibfnamefont{V.}~\bibnamefont{Terras}},
  \bibinfo{journal}{Nucl. Phys. B} \textbf{\bibinfo{volume}{654}},
  \bibinfo{pages}{323} (\bibinfo{year}{2003}).

\bibitem[{\citenamefont{Sirker and Bortz}(2006)}]{sir06}
\bibinfo{author}{\bibfnamefont{J.}~\bibnamefont{Sirker}} \bibnamefont{and}
  \bibinfo{author}{\bibfnamefont{M.}~\bibnamefont{Bortz}}, \bibinfo{journal}{J.
  Stat. Mech.} p. \bibinfo{pages}{P01007} (\bibinfo{year}{2006}).

\bibitem[{\citenamefont{Deguchi et~al.}(2001)\citenamefont{Deguchi, Fabricius,
  and McCoy}}]{deg01}
\bibinfo{author}{\bibfnamefont{T.}~\bibnamefont{Deguchi}},
  \bibinfo{author}{\bibfnamefont{K.}~\bibnamefont{Fabricius}},
  \bibnamefont{and} \bibinfo{author}{\bibfnamefont{B.~M.} \bibnamefont{McCoy}},
  \bibinfo{journal}{J. Stat. Phys.} \textbf{\bibinfo{volume}{102}},
  \bibinfo{pages}{701} (\bibinfo{year}{2001}).

\bibitem[{\citenamefont{Braak and Andrei}(2001)}]{bra01}
\bibinfo{author}{\bibfnamefont{D.}~\bibnamefont{Braak}} \bibnamefont{and}
  \bibinfo{author}{\bibfnamefont{N.}~\bibnamefont{Andrei}},
  \bibinfo{journal}{J. Stat. Phys.} \textbf{\bibinfo{volume}{105}},
  \bibinfo{pages}{677} (\bibinfo{year}{2001}).

\bibitem[{\citenamefont{Fabricius and McCoy}(2001{\natexlab{a}})}]{fab01a}
\bibinfo{author}{\bibfnamefont{K.}~\bibnamefont{Fabricius}} \bibnamefont{and}
  \bibinfo{author}{\bibfnamefont{B.~M.} \bibnamefont{McCoy}},
  \bibinfo{journal}{J. Stat. Phys.} \textbf{\bibinfo{volume}{103}},
  \bibinfo{pages}{647} (\bibinfo{year}{2001}{\natexlab{a}}).

\bibitem[{\citenamefont{Fabricius and McCoy}(2001{\natexlab{b}})}]{fab01b}
\bibinfo{author}{\bibfnamefont{K.}~\bibnamefont{Fabricius}} \bibnamefont{and}
  \bibinfo{author}{\bibfnamefont{B.~M.} \bibnamefont{McCoy}},
  \bibinfo{journal}{J. Stat. Phys.} \textbf{\bibinfo{volume}{104}},
  \bibinfo{pages}{573} (\bibinfo{year}{2001}{\natexlab{b}}).

\bibitem[{\citenamefont{Nomura}(1993)}]{nom93}
\bibinfo{author}{\bibfnamefont{K.}~\bibnamefont{Nomura}},
  \bibinfo{journal}{Phys. Rev. B} p. \bibinfo{pages}{16814}
  (\bibinfo{year}{1993}).

\bibitem[{\citenamefont{Haldane}(1991)}]{hal91}
\bibinfo{author}{\bibfnamefont{F.~D.~M.} \bibnamefont{Haldane}},
  \bibinfo{journal}{Phys. Rev. Lett.} \textbf{\bibinfo{volume}{67}},
  \bibinfo{pages}{937} (\bibinfo{year}{1991}).

\bibitem[{\citenamefont{Arikawa et~al.}(2006)\citenamefont{Arikawa, Karbach,
  M\"uller, and Wiele}}]{ari06}
\bibinfo{author}{\bibfnamefont{M.}~\bibnamefont{Arikawa}},
  \bibinfo{author}{\bibfnamefont{M.}~\bibnamefont{Karbach}},
  \bibinfo{author}{\bibfnamefont{G.}~\bibnamefont{M\"uller}}, \bibnamefont{and}
  \bibinfo{author}{\bibfnamefont{K.}~\bibnamefont{Wiele}}, \bibinfo{journal}{J.
  Phys. A} \textbf{\bibinfo{volume}{39}}, \bibinfo{pages}{10623}
  (\bibinfo{year}{2006}).

\bibitem[{\citenamefont{Karbach and M\"uller}(2000)}]{kar00}
\bibinfo{author}{\bibfnamefont{M.}~\bibnamefont{Karbach}} \bibnamefont{and}
  \bibinfo{author}{\bibfnamefont{G.}~\bibnamefont{M\"uller}},
  \bibinfo{journal}{Phys. Rev. B} \textbf{\bibinfo{volume}{62}},
  \bibinfo{pages}{14871} (\bibinfo{year}{2000}).

\bibitem[{\citenamefont{Bortz and Stolze}(2007)}]{bor07}
\bibinfo{author}{\bibfnamefont{M.}~\bibnamefont{Bortz}} \bibnamefont{and}
  \bibinfo{author}{\bibfnamefont{J.}~\bibnamefont{Stolze}},
  \bibinfo{journal}{Phys. Rev. B} \textbf{\bibinfo{volume}{76}},
  \bibinfo{pages}{014304} (\bibinfo{year}{2007}).

\bibitem[{\citenamefont{Faribault et~al.}(2008)\citenamefont{Faribault,
  Calabrese, and Caux}}]{far08}
\bibinfo{author}{\bibfnamefont{A.}~\bibnamefont{Faribault}},
  \bibinfo{author}{\bibfnamefont{P.}~\bibnamefont{Calabrese}},
  \bibnamefont{and} \bibinfo{author}{\bibfnamefont{J.-S.} \bibnamefont{Caux}},
  \bibinfo{journal}{J. Stat. Mech.} p. \bibinfo{pages}{P03018}
  (\bibinfo{year}{2009}).

\bibitem[{\citenamefont{Kitanine et~al.}(2009)\citenamefont{Kitanine,
  Kozlowski, Maillet, Slavnov, and Terras}}]{kit09}
\bibinfo{author}{\bibfnamefont{N.}~\bibnamefont{Kitanine}},
  \bibinfo{author}{\bibfnamefont{K.}~\bibnamefont{Kozlowski}},
  \bibinfo{author}{\bibfnamefont{J.}~\bibnamefont{Maillet}},
  \bibinfo{author}{\bibfnamefont{N.}~\bibnamefont{Slavnov}}, \bibnamefont{and}
  \bibinfo{author}{\bibfnamefont{V.}~\bibnamefont{Terras}},
  \bibinfo{journal}{arXiv:0903.2916}  (\bibinfo{year}{2009}).

\end{thebibliography}
\end{document}